\documentclass[reprint,superscriptaddress,aps,prx,nofootinbib,longbibliography]{revtex4-2}
\usepackage{LatexStyle}
\usepackage{adjustbox}
\usepackage{media9} 
\usepackage{multimedia}
\usepackage{silence}
\usepackage{standalone}
\usepackage[english]{babel}
\usepackage{subfiles}
\usepackage{hyperref}
\hypersetup{
    unicode=true,
    pdfusetitle,
    bookmarks=true,
    bookmarksnumbered=true,
    bookmarksopen=false,
    breaklinks=false,
    pdfborder={0 0 0},
    pdfborderstyle={},
    backref=false,
    colorlinks=false,
    pdftitle={Effective Quantization of Lossy Nonlinear Epsilon-Near-Zero Media}
}
\usepackage[dvipsnames]{xcolor}
\begin{document}
\ActivateWarningFilters[pdftoc]

\title{Effective Quantization of Lossy Nonlinear Epsilon-Near-Zero Media}
\author{Avishi Poddar}
\email{avishipoddar@g.harvard.edu}
\affiliation{Department of Physics, Harvard University, Cambridge, Massachusetts 02138, USA}
\author{Jonas von Milczewski}
\affiliation{Department of Physics, Harvard University, Cambridge, Massachusetts 02138, USA}
\author{Durdu O. Guney}
\affiliation{Department of Electrical and Computer Engineering, Michigan Technological University, Houghton, MI 49931, USA}
\author{Sahin K. \"Ozdemir}
\affiliation{Department of Electrical and Computer Engineering, St. Louis University, St. Louis, MO USA}
\author{Susanne F. Yelin}
\affiliation{Department of Physics, Harvard University, Cambridge, Massachusetts 02138, USA}
\date{\today}

\begin{abstract}
In epsilon-near-zero (ENZ) materials, the vanishing real linear permittivity results in the leading-order contribution to the displacement field being nonlinear in the electric field, making conventional canonical quantization approaches highly non-trivial. Existing treatments generally quantize the linear modes first and introduce nonlinear interactions subsequently, an ordering that becomes inadequate in the ENZ regime. Here, we provide, to our knowledge, the first effective single-excitation quantization in which the near-zero linear response, leading nonlinearity, and loss jointly determine the elementary excitation. Using a solvable microscopic atomic system as a theoretical scaffold, we find that this excitation is a polariton: a dressed quasiparticle with partly field and partly material excitations whose coefficients can be parameterized by macroscopic susceptibilities. While this work considers a leading-order $\chi^{(3)}$ nonlinearity, the framework could be generalized to include other nonlinear corrections, including a controllable second-order susceptibility $\chi^{(2)}$ and systematic higher-order contributions $\chi^{(n)}$, $n>2$. Our findings pave the way toward practical applications in quantum photonics, such as single-photon non-demolition detection, by leveraging the strong nonlinearities intrinsic to zero-index materials.
\end{abstract}

\maketitle

\section{Introduction}

Epsilon-near-zero (ENZ) materials, such as indium tin oxide (ITO), aluminum-doped zinc oxide (AZO), and cadmium oxide (CdO), have a unique ability to modify both the linear and nonlinear propagation characteristics of electromagnetic waves near frequencies at which the real part of their dielectric permittivity approaches zero \cite{2019NatRM}. Field enhancement, slow-light effects, and the small linear index can strongly enhance nonlinear optical responses \cite{alam_large_2016}. In the low-loss limit, the corresponding refractive index experienced by light of a particular frequency becomes small, resulting in increased wavelength and phase velocity \cite{2019NatRM,Kinsey2021}. Consequently, the light field acquires no spatial phase over macroscopic distances \cite{2019NatRM,2017NaPho..11..149L}. 
Based on these novel properties, near-zero-index media provide a promising avenue for various applications, including supercoupling through narrow channels and sharp bends \cite{PhysRevE.78.016604}, geometry-independent resonant structures \cite{liberal_geometry-invariant_2016}, enhanced long-range interactions between spatially separated emitters and superradiance \cite{Mahmoud:17,mello2025longrange,mello2022extended}, and compact nonlinear and integrated photonic devices \cite{2019NatRM,Kinsey2021}. 
More recently, the large Kerr response available in ENZ nanostructures has motivated proposals for single-photon nonlinear optics, including quantum non-demolition (QND) detection \cite{LDN2025}. While this makes ENZ materials a compelling platform in nonlinear optics, it presents a nontrivial challenge for quantization. Previous approaches quantize the linear excitations and subsequently add the nonlinearity as an interaction. Here, because the linear response vanishes, the nonlinear response participates in determining what the elementary excitation itself is. 

Standard linear quantization techniques involve starting with a classical Lagrangian density whose equations of motion reproduce the macroscopic Maxwell equations. 
Using the constitutive relation linking the displacement field $\mathbf{D}$ and the electric field $\mathbf{E}$, $\mathbf{D}=\epsilon_0\mathbf{E}+\mathbf{P}$, where $\mathbf{P}$ is the induced polarization, the Hamiltonian is expressed in terms of the canonical field variables. The theory is then quantized by promoting the classical variables to operators by imposing equal-time commutation relations. In microscopic descriptions, diagonalization of the coupled light-matter Hamiltonian leads to collective matter-field excitations (polaritons) \cite{Hopfield}.

This procedure is well-established for linear dielectrics. In the Huttner-Barnett construction, for instance, the electromagnetic field is coupled to a polarization field and a reservoir, providing a full canonical quantization of a linear dispersive and absorptive dielectric while respecting Kramers-Kronig relations \cite{HuttnerBarnett}. Related macroscopic-QED formulations instead take the complex dielectric function as input and express the quantized electromagnetic field in terms of the corresponding Green's tensor and Langevin noise operators \cite{gruner_green-function_1996,knoll2003qeddispersingabsorbingmedia}. Path-integral approaches based on the microscopic Hopfield-type dielectric models have also been developed \cite{Bechler}.

Extensions of dielectric quantization to incorporate nonlinearities have also been considered in certain contexts \cite{Hillery_2009,HilleryMlodinow,Drummond,DrummondHillery}. Within macroscopic QED, nonlinear interaction Hamiltonians and associated nonlinear noise polarization operators have been constructed from the quantized linear theory \cite{PhysRevLett.96.073601}. Path integral approaches provide a formal description of nonlinear optical media, although practical calculations are usually performed under approximations such as the undepleted pump approximation \cite{difallah_path-integral_2019}. More recent mesoscopic approaches introduce nonlinear matter interactions explicitly and formulate their effects through a perturbative diagrammatic expansion \cite{Scheel2026}. At the classical field-theory level, fully non-perturbative descriptions of the optical nonlinearity of ENZ materials have also been recently developed \cite{tamashevich_field_2024}. While these works have developed increasingly general treatments of nonlinear dielectric quantization, our work combines the central ingredients relevant to ENZ --- near-zero real linear permittivity, appreciable $\chi^{(3)}$, and loss --- in a single-excitation construction, to provide a susceptibility-parameterized polariton description of a lossy nonlinear ENZ medium.

Developing such a theory for a realistic (lossy) medium that exhibits a near-zero index and strong nonlinearity remains largely unexplored due to a number of reasons. In a low-loss ENZ material, the vanishing real linear electric permittivity means the dominant contribution to $\mathbf{D}$ is nonlinear in the electric field; this means the nonlinear polarization is no longer a small perturbative correction to the linear response. Due to this inherently nonlinear relationship between the field and its displacement, the typical power-series method to obtain $\mathbf{E}\pare{\mathbf{D}}$ ceases to provide a controlled expansion. Correspondingly, the standard Hamiltonian description of the electromagnetic field as a collection of harmonic oscillators which arises from a quadratic field Hamiltonian breaks down in the ENZ regime due to the inherently anharmonic nature of the system. Despite formal nonlinear quantization schemes, the challenge therefore remains to obtain a useful nonperturbative set of quantum excitations that accurately capture the regime in which the nonlinear response is dominant, with a nominal linear response. Overcoming this challenge is essential not only from a fundamental perspective, but also from a practical one, enabling the accurate design and performance prediction of quantum optical systems involving ENZ media.

In this work, we circumvent the aforementioned challenges with conventional quantization schemes by leveraging a solvable microscopic model to construct a phenomenological, polariton-based quantized description of a nonlinear, lossy ENZ material. We employ a solvable atomic model as a theoretical scaffold to derive a self-consistent mathematical form of the single-excitation quasiparticle operators and subsequently connect their parameters to the macroscopic response of the ENZ medium. In this way, the modification of the elementary excitations by the nonlinear response is directly incorporated into the definition of the dressed quasiparticle. Our approach provides a complementary perspective to recent Green's tensor quantization approaches to nonlinear QED of ENZ nanostructures \cite{LDN2025}. In that formulation, the linear theory is first quantized using a Langevin-noise Green's tensor approach, and the Kerr nonlinearity is subsequently introduced through an interaction Hamiltonian among those linear dressed excitations. Here, by contrast, our objective is to construct an effective single-excitation quasiparticle operator whose field-matter composition is determined self-consistently by the nonlinear material response.

\section{General Approach}

We first present an overview of our approach, followed by results in the next section. Our construction proceeds in three steps. First, we choose a solvable microscopic open-system model whose steady-state response reproduces prescribed linear and nonlinear susceptibilities near the ENZ frequency. Second, we invert this response to express the microscopic parameters in terms of the macroscopic susceptibilities. Finally, we diagonalize the resulting non-Hermitian model in the single-excitation sector. The resulting polariton operators, therefore, provide an effective quantum description whose field–matter composition is determined by the nonlinear ENZ response.

A typical ENZ material is characterized by a vanishing linear permittivity $\pare{\Re\pare{\chi^{\pare{1}}}\to -1}$, low optical losses $\pare{\text{small }\Im\pare{\chi^{\pare{1}}}}$, and an appreciable nonlinearity $\pare{\abs{\chi^{\pare{3}}}>0}$. To relate microscopic parameters of a quantized theory to macroscopic observables, we first establish a correspondence between the macroscopic susceptibilities of a nonlinear ENZ material and the microscopic parameters of a solvable atomic system within a semiclassical framework. To this end, we introduce a minimal atomic model capable of reproducing this response and obtain the macroscopic polarization induced by the field which depends on steady-state atomic coherences. Within this chosen microscopic model, the parameters can be inverted in terms of the target macroscopic susceptibilities. Our microscopic model does not span the full space of nonlinear ENZ responses but provides a representative framework that can be applied to ENZ media in a certain regime.

\begin{figure} [!t]
    \centering
    \includegraphics[width=\linewidth]{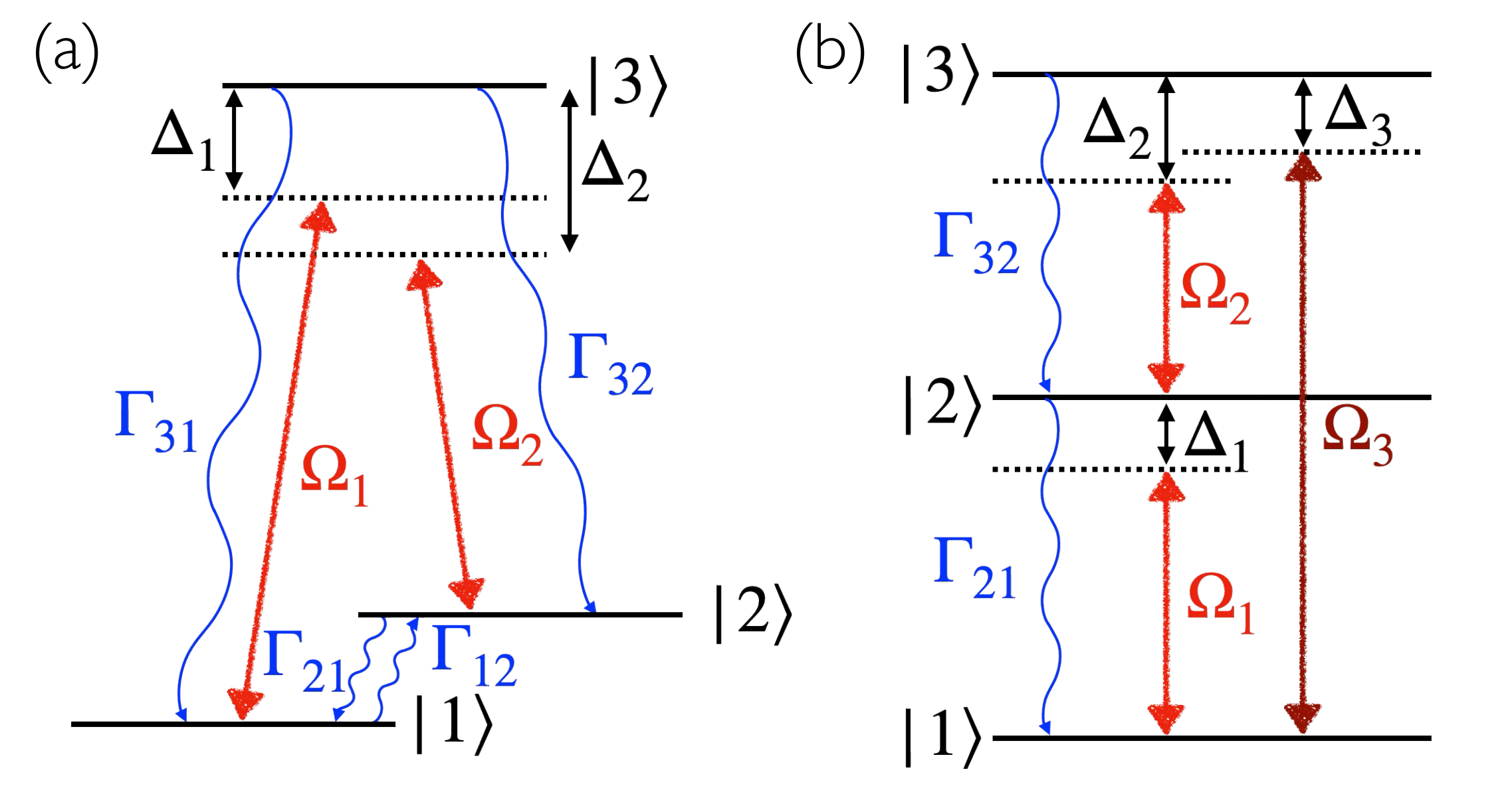}
    \caption{(a) The $\Lambda$-type atomic system used to model an ENZ response. Electric fields couple to the $\ket{1}-\ket{3}$ and the $\ket{2}-\ket{3}$ atomic transitions with Rabi frequencies $\Omega_j=\mu_{3j}\langle\hat{\mathcal{E}}_j\rangle/\hbar$ and detunings $\Delta_j$. The excited state $\ket{3}$ decays to the states $\ket{1}$ and $\ket{2}$ with rates $\Gamma_{31}$ and $\Gamma_{32}$. (b) An analogous atomic model that could be used to generate a second-order steady-state coherence term. One electric field at frequency $\omega$ couples the $\ket{1}-\ket{2}$ and $\ket{2}-\ket{3}$ transitions with Rabi frequencies $\Omega_1$ and $\Omega_2$ and detunings $\Delta_1$ and $\Delta_2$. A second electric field at frequency $2\omega$ couples the $\ket{1}-\ket{3}$ transition with Rabi frequency $\Omega_3$ and detuning $\Delta_3$.}
    \label{fig:atomicmodel}
\end{figure}

We model the material as an ensemble of $N$ identical, non-interacting atoms confined to a volume $V$ with number density $N/V$. We perform the calculations for a single atom and scale the resulting macroscopic observables by the number density. Without loss of generality, we consider the field(s) as a monochromatic plane wave $\hat{\mathcal{E}_j}$ propagating in the $\hat{z}-$direction with a slowly-varying field amplitude and a field frequency $\nu_j$. The field operator $\hat{\mathcal{E}}_j$ is proportional to a bosonic annihilation operator $\hat{a}_j$, $\hat{\mathcal{E}}_j\left(z,t\right) = \sqrt{\hbar \nu_j/2\epsilon_0 V} \hat{a}_j\left(z,t\right)$. The coupling strength $g_j$ assumed to be real is given by $g_j = \sqrt{\nu_j/\pare{2\hbar\epsilon_0V}}\mu_{3j}$. A minimal atomic model that addresses the requirements of a typical ENZ material is that of a $\Lambda-$type system, with two ground states $\ket{1}$ and $\ket{2}$ and an excited state $\ket{3}$, shown in \figref{fig:atomicmodel}a. The $\ket{1}-\ket{3}$ and $\ket{2}-\ket{3}$ are driven by electric fields with Rabi frequencies $\Omega_1$ and $\Omega_2$ and detunings $\Delta_1$ and $\Delta_2$ respectively. Additionally, the excited state $\ket{3}$ can decay to the states $\ket{1}$ and $\ket{2}$ at rates $\Gamma_{31}$ and $\Gamma_{32}$. There can also exist ground-state relaxation between the states $\ket{1}$ and $\ket{2}$ at rates $\Gamma_{12}$ and $\Gamma_{21}$. This model is minimal in the sense that it has the minimum necessary levels to achieve the required tunability of the linear and nonlinear responses; however, this choice is not unique, and other microscopic models could yield equivalent macroscopic results.

The interaction of a single atom with the field, under the rotating wave approximation and in a frame rotating at frequency $\nu$, is described by the following Hamiltonian:
\be\label{eq:H_SC}
    \frac{\hat{H}}{\hbar} = \pare{\Delta_1-\Delta_2}\hat{\sigma}_{22} + \Delta_1\hat{\sigma}_{33} - \pare{g_1 \hat{\sigma}_{31} \hat{a}_1 + g_2 \hat{\sigma}_{32}\hat{a}_2 + \text{h.c.}},
\ee
where $\Delta_j$ are the detunings for the $\ket{3}-\ket{j}$ transitions, $\Delta_j=\omega_{3}-\omega_{j}-\nu_j$. The steady-state of the model can be easily obtained in the mean field limit, $\hat{\mathcal{E}}_j\to\mathcal{E}_j\equiv\langle\hat{\mathcal{E}}_j\rangle$. Assuming a Markovian system, its dynamics are governed by the Lindblad master equation $\dot{\hat{\rho}} =\mathcal{L}\left[\hat{\rho}\right]$ which can be solved to obtain the single-atom finite-drive steady-state density matrix $\hat{\rho}_{\text{ss}}$ such that $\mathcal{L}\left[\hat{\rho}_{\text{ss}}\right]=0$. The macroscopic polarization amplitude is then determined by the ensemble-averaged dipole moment, which depends on the steady-state atomic coherences. This procedure establishes a direct mapping from microscopic atomic parameters like atomic detunings and decay rates, to the macroscopic material observables like susceptibilities, $\chi^{\pare{1}}$ and $\chi^{\pare{3}}$ \cite{Feizpour_2015}.

We then proceed to a fully quantized version of this atomic model. Since we consider non-interacting atoms, we perform all relevant calculations in the projected single-atom subspace. The microscopic quantization of the system is described by polariton creation and annihilation operators $\hat{P}_j^{\pare{\dag}}$ with energy $E_j$. These operators define the normal modes of the system and can be found by diagonalizing the full time evolution within the single excitation subspace. In the low-excitation regime, the dynamics in this subspace are captured by the non-Hermitian effective Hamiltonian 
\be\label{eq:Heff}
    \frac{\hat{H}_{\text{eff}}}{\hbar}\equiv \frac{\hat{H}}{\hbar}-\frac{i\Gamma_{12}}{2}\sigop_{11}-\frac{i\Gamma_{21}}{2}\sigop_{22}-\frac{i\pare{\Gamma_{31}+\Gamma_{32}}}{2}\sigop_{33},
\ee
allowing us to find the normal modes by solving the following equation,
\begin{equation}
\begin{split}\label{eq:PolEv}
    \frac{d\hat{P}_j^\dag}{dt} &= -\frac{i}{\hbar}\left(\hat{P}_j^\dag\hat{H}_{\text{eff}} 
    -\hat{H}_{\text{eff}}^\dag \hat{P}_j^\dag\right) = -iE_j\hat{P}_j^\dag,
\end{split}
\end{equation}
neglecting the jump terms in this subspace. Here, $E_j=k_j+i\kappa_j$ is the complex eigenfrequency of the polariton, where $k_j$ represents its frequency and $\kappa_j\pare{<0}$ its decay rate.

This system admits two types of polariton creation operators, which we label as $\hat{A}^\dag_j$ and $\hat{B}^\dag_j$, corresponding to creating an excitation by their respective action on the zero-excitation states $\ket{1}$ and $\ket{2}$. These operators describe a single excitation quasiparticle that is part photonic and part atomic. The relative weights of the photonic and atomic components are determined from the right eigenvectors of $\hat{H}_{\text{eff}}^\dag$ projected onto the single excitation subspace. For each type of polariton operator, there are three distinct solutions $\pare{j=1,2,3}$, each with a specific complex energy. The energies $E_j$ and coefficients $c_{ij}$ are functions of the system parameters. The two types of polariton operators $\hat{A}_j^\dag$ and $\hat{B}_j^\dag$ are constructed from the same coefficients $c_{ij}$ and differ only in the ground state from which they are built. Depending on the ground state the system is initialized in, one must consider $\hat{A}_j^\dag$ or $\hat{B}_j^\dag$ or their relevant combination.

Using the previously established mean-field mapping, one can obtain $E_j$ and $c_{ij}$ in terms of macroscopic material parameters. This allows a detailed analysis of the nature of the quantized quasiparticles of the system as a function of their susceptibilities. Additionally, such a description incorporates losses that are inherent to realistic materials. While in ideal lossless systems, the polariton operators obey canonical bosonic commutation relations, the presence of loss results in polaritons that are no longer perfectly bosonic, reflecting the fact that the elementary excitations are quasiparticles with finite lifetimes.

Although this work focuses on a $\chi^{\pare{3}}$ nonlinearity, \figref{fig:atomicmodel}b illustrates an analogous ladder-type atomic model capable of generating a second-order coherence.
In this model, there are two electric fields at frequencies $\omega$ and $2\omega$, with the first field coupling the $\ket{1}-\ket{2}$ and $\ket{2}-\ket{3}$ transitions which have approximately similar energy gaps and the second coupling the $\ket{1}-\ket{3}$ transition. In solid-state materials, symmetry-breaking mechanisms yield a response akin to the $\chi^{(2)}$ response generated by the term proportional to the second-order coefficient of $\rho_{31}^{\text{ss}}$ for non-zero $\mu_{31}$ \cite{Boyd}. One can then follow a similar procedure as described for the $\Lambda$-type atomic model to establish a mapping between the microscopic and macroscopic parameters and then establish the corresponding quantization.

\section{Results}

We first present the case that a single field drives both atomic transitions $\ket{1}-\ket{3}$ and $\ket{2}-\ket{3}$. From a macroscopic perspective, this means that the signal and probe fields are indistinguishable. The single-field case captures the essential structure of the susceptibility mapping and the polariton construction in its simplest analytical form. We then show that this construction readily extends to the case of distinguishable signal and probe fields, where the nonlinearity contains distinct self- and cross-terms.

\begin{figure}
    \centering
    \includegraphics[width=\linewidth]{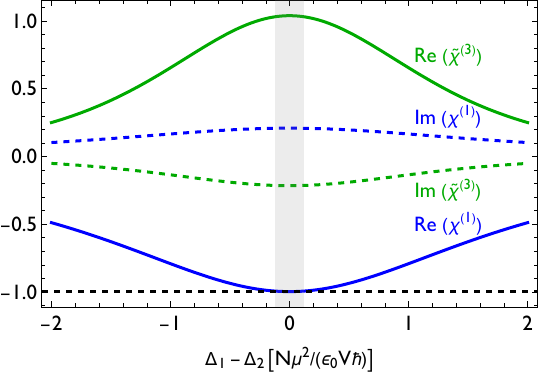}
    \caption{Example of the ENZ response generated by a $\Lambda$-system coupled to a single field for $\Gamma=0.2N \mu^2/(\epsilon_0 V \hbar)$ and $\Delta_1+\Delta_2=-1.92N \mu^2/(\epsilon_0 V \hbar)$.
    The solid (dashed) blue and green lines represent the real (imaginary) parts of $\chi^{\pare{1}}$ and $\tilde{\chi}^{\pare{3}}\equiv \chi^{\pare{3}}\pare{N\mu/(\epsilon_0 V)}^2$ respectively. The region corresponding to an ENZ response is shaded.}
    \label{fig:ENZresponse_singlefield}
\end{figure}

In the case that there is only one field coupling both atomic transitions, $\mathcal{E}_1=\mathcal{E}_2\equiv\mathcal{E}$. The macroscopic polarization amplitude is then determined by the ensemble-averaged dipole moment, which depends on the steady-state atomic coherences as follows
\be\label{eq:Pol_SC}
\begin{split}
    \mathcal{P} &
    =  \frac{N}{V}\pare{
    \mu_{31}\rho_{31}^{\text{ss}}
    + \mu_{32}\rho_{32}^{\text{ss}}}\approx\epsilon_0\pare{\chi^{\pare{1}}\mathcal{E}+\chi^{\pare{3}}\abs{\mathcal{E}}^2\mathcal{E}+...},
\end{split}
\ee
where $\rho_{ji}^{\text{ss}} = \Tr\pare{\ket{i}\bra{j}\hat{\rho}_{\text{ss}}}$. For simplicity, we assume that $\Gamma_{12}=\Gamma_{21}=0$, $\Gamma_{31}=\Gamma_{32}\equiv\Gamma\geq0$, $\mu_{31} = \mu_{32} \equiv \mu \in \mathbb{R}$, and $g_1=g_2\equiv g$. 
By expanding the resulting expression for $\mathcal{P}$ in powers of the electric field amplitude $\mathcal{E}$, we can identify the terms corresponding to the linear $\pare{\chi^{(1)}}$ and third-order $\pare{\chi^{(3)}}$ susceptibilities, which are given by
\be\label{eq:rechi1}
    \Re\pare{\chi^{(1)}} = \frac{N}{V\epsilon_0}\frac{(\Delta_1 + \Delta_2) \mu^2}{(2\Gamma^2 + \Delta_2^2 + \Delta_1^2) \hbar},
\ee
\be\label{eq:imchi1}
    \Im\pare{\chi^{(1)}} = \frac{2\Re\pare{\chi^{(1)}}\Gamma}{\Delta_1+\Delta_2},
\ee
\be\label{eq:rechi3}
    \Re\pare{\chi^{(3)}} = -\frac{2V\epsilon_0}{N\hbar} \frac{\pare{\Re\pare{\chi^{(1)}}}^2}{\pare{\Delta_1 + \Delta_2}},
\ee
\be\label{eq:imchi3}
    \Im\pare{\chi^{(3)}} =\frac{2\Re\pare{\chi^{(3)}}\Gamma}{\Delta_1+\Delta_2}.
\ee
From the above equations, it is evident that when $\Re\pare{\chi^{(1)}}\approx -1$, $\Delta_1+\Delta_2<0$, which automatically gives $\Re\pare{\chi^{(3)}}>0$, as is the case with typical ENZ materials. We note that the fixed ratio between the real and imaginary parts of the susceptibility is not a universal constraint; it is a consequence of the fact that there is one common driving field for both the linear and third-order responses in our model. Since the absorptive and dispersive parts originate from the same damped resonance, they cannot be tuned independently within this minimal model. 

At a fixed operating frequency, the real and imaginary parts of $\chi^{\pare{1}}$ and $\chi^{\pare{3}}\spare{\pare{\epsilon_0 V/(N\mu)}^2}$ determine the total detuning $\Delta_1+\Delta_2$, while the relative detuning $\Delta_1-\Delta_2$ remains free. In a real material, this parameter would be fixed by additional spectral information, such as the slope of $\chi^{\pare{1}}$ near the ENZ point or the ENZ bandwidth.
\figref{fig:ENZresponse_singlefield} shows the dependence of the susceptibilities on this relative detuning parameter, for a fixed decay rate $\Gamma=0.2N \mu^2/(\epsilon_0 V \hbar)$. For a  certain range of small relative detuning values, the desired ENZ response is obtained, as indicated by the shaded region. 

This mean-field calculation establishes a direct mapping from the microscopic atomic parameters $\pare{\Delta_j,\Gamma,\mu}$ to the macroscopic observables $\pare{\chi^{\pare{1}}, \chi^{\pare{3}},\delta}$, where $\delta$ is a relative detuning parameter $\delta\equiv \Delta_1-\Delta_2$. 
As a result, the microscopic atomic parameters may be expressed entirely in terms of the measurable material properties,
\be\label{eq:Gamma_final}
    \Gamma = -\frac{V\epsilon_0}{N\hbar}\frac{\Re\pare{\chi^{(1)}}\Im\pare{\chi^{(1)}}}{\Re\pare{\chi^{(3)}}},
\ee
\be\label{eq:delta1_final}
    \Delta_1 = \frac{1}{2}\pare{\delta-2\frac{V\epsilon_0}{N\hbar}\frac{\Re\pare{\chi^{(1)}}^2}{\Re\pare{\chi^{(3)}}}},
\ee
\be\label{eq:delta2_final}
    \Delta_2 = -\frac{1}{2}\pare{\delta+2\frac{V\epsilon_0}{N\hbar}\frac{\Re\pare{\chi^{(1)}}^2}{\Re\pare{\chi^{(3)}}}},
\ee
\be\label{eq:mu_final}
    \frac{\mu^2}{\hbar^2} = \pare{\frac{V\epsilon_0}{N\hbar}}^2
    \frac{\pare{-\Re\pare{\chi^{(1)}}}}{\Re\pare{\chi^{(3)}}}
    \abs{\chi^{(1)}}^2 
    +\frac{\delta^2}{4}\frac{\Re\pare{\chi^{(3)}}}{\pare{-\Re\pare{\chi^{(1)}}}}.
\ee
This inversion is the key result of the semiclassical analysis and forms the foundation for building the phenomenological quantized description.

The microscopic quantization of the system is described by the action of two types of polariton operators, 
$\hat{A}^\dag_j$ and $\hat{B}^\dag_j$, corresponding to creating an excitation by their respective action on the zero-excitation states $\ket{1}$ and $\ket{2}$. They have the form:
\be\label{eq:Adag}
    \hat{A}_j^\dag = c_{1j} \hat{a}^\dag\hat{\sigma}_{11} + c_{2j} \hat{a}^\dag\hat{\sigma}_{21} + c_{3j} \hat{\sigma}_{31},
\ee
\be\label{eq:Bdag}
    \hat{B}_j^\dag = c_{1j} \hat{a}^\dag\hat{\sigma}_{12} + c_{2j} \hat{a}^\dag\hat{\sigma}_{22} + c_{3j} \hat{\sigma}_{32}.
\ee
The two operators $\hat{A}^\dag_j$ and $\hat{B}^\dag_j$ create the same dressed single-excitation eigenvector from different reference states and hence share the same coefficients $c_{ij}$, the exact expression for which can be found in \appref{app:PolaritonCoeffs}. Their corresponding energies differ by the relative detuning $\delta$. We refer to the energy associated with $\hat{A}_j^\dag$ as $E_j$ and correct by the relative detuning to obtain the energy of $\hat{B}_j^\dag$.

\begin{figure*}
     \centering
    \includegraphics[width=0.9\textwidth]{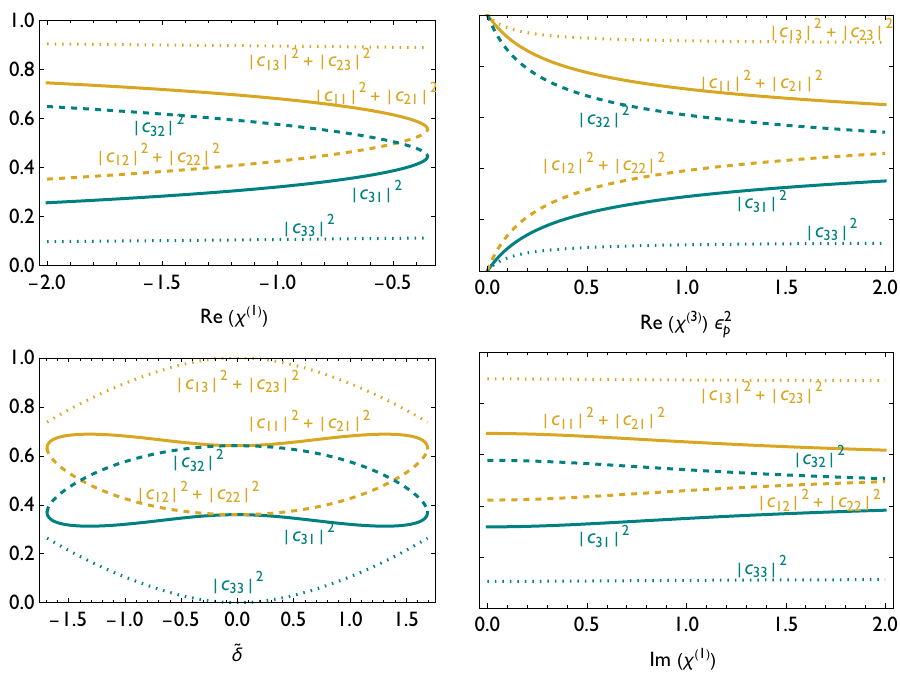}
    \caption{Relative weights of the photonic $\pare{\abs{c_{1j}}^2+\abs{c_{2j}}^2}$ and atomic $\pare{\abs{c_{3j}}^2}$ parts of the polariton operators. The representative values of parameters chosen are $\text{Re}\pare{\chi^{(1)}}=-1$, $\text{Re}\pare{\chi^{(3)}}=1.4/\epsilon_p^2$, $\tilde{\delta}=1$, and $\text{Im}\pare{\chi^{(1)}}=0.2$. In each subplot, one of these values is varied, keeping the other three fixed.}
    \label{fig:polaritonweights_singlefield}
\end{figure*}

Using the previously established mean-field mapping (Eqs.~(\ref{eq:Gamma_final})-(\ref{eq:mu_final})), $E_j$ and $c_{ij}$ can be re-expressed solely in terms of the linear and third-order susceptibilities, $\chi^{(1)}$ and $\chi^{(3)}$, and a dimensionless detuning parameter $\tilde{\delta}\equiv\delta/g$.
\figref{fig:polaritonweights_singlefield} shows the relative weights of the photonic $\pare{\abs{c_{1j}}^2+\abs{c_{2j}}^2}$ and atomic $\pare{\abs{c_{3j}}^2}$ parts of the polariton operators. We express $\chi^{(3)}$ in units of the inverse square of the field strength for a single photon $\epsilon_p\equiv\hbar g/\mu$. Each set of values for $\chi^{(1)}$, $\chi^{(3)}$, and $\tilde{\delta}$ corresponds to a particular relative weight of the atomic and photonic components, thus quantifying the exact constitution of the polariton quasiparticle operator representing one excitation of that system.

\begin{figure}
    \centering
    \includegraphics[width=\linewidth]{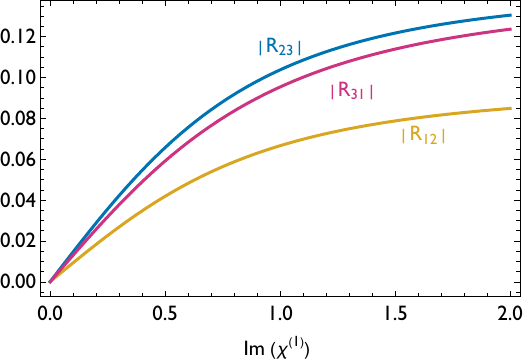}
    \caption{Deviation from bosonic commutation of the polariton operators as a function of the imaginary part of the linear susceptibility $\text{Im} \spare{\chi^{(1)}}$ which quantifies loss in the system for $\text{Re}\pare{\chi^{(1)}}=-1$, $\text{Re}\pare{\chi^{(3)}}=1.4/\mathcal{E}_p^2$, and $\tilde{\delta}=1.5$.}
    \label{fig:bosdev}
\end{figure}

In an ideal lossless system $\pare{\Gamma=0}$, the effective Hamiltonian $\hat{H}_{\text{eff}}$ would be Hermitian. Consequently, its eigenstates (which determine the polaritonic coefficients) are perfectly orthonormal. This ensures that the polariton operators obey canonical bosonic commutation relations, $\langle 1 | \left[ \hat{A}_i, \hat{A}_j^\dag \right]|1 \rangle = \langle 2|\left[\hat{B}_i,\hat{B}_j^\dag\right]|2 \rangle = \delta_{ij}$. However, in realistic systems with loss, the decay rate does not vanish $\Gamma\neq 0$, which renders $\hat{H}_{\text{eff}}$ non-Hermitian. A fundamental consequence of this is that its eigenstates, though still normalized, are not perfectly orthogonal; the polaritons no longer behave as ideal bosons but their commutation relation is modified to $\langle 1|\left[\hat{A}_i,\hat{A}_j^\dag\right]|1 \rangle = \langle 2|\left[\hat{B}_i,\hat{B}_j^\dag\right]|2 \rangle = \delta_{ij}+\pare{1-\delta_{ij}} 2i\Gamma c_{3i}^*c_{3j}/\pare{\pare{E_i}^*-E_j}\equiv R_{ij}$. This deviation from bosonic nature depends on the value of the imaginary part of the linear susceptibility, as shown in \figref{fig:bosdev}; as $\text{Im} \spare{\chi^{(1)}}\to 0$, the polaritons approach perfect bosonic behavior. The off-diagonal terms $\pare{i\neq j}$ are a direct measure of the dissipative coupling between the different polariton modes and reflect the fact that the elementary excitations are quasiparticles with finite lifetimes.

We can now use this to rewrite the effective Hamiltonian (conjugate transpose) as follows:
\be\begin{split}\label{eq:heff_nh}
    \frac{\hat{H}_{\text{eff}}^\dag}{\hbar} &= \delta\ket{2,0}\bra{2,0} + \sum\limits_{j,k = 1}^3 M_{jk}\hat{P}_j^\dag \hat{P}_k,
\end{split}\ee
where $\hat{P}_j$ can be any (normalized) superposition of $\hat{A}_j$ and $\hat{B}_j$, $M_{jk}$ are elements of the matrix $\mathbf{M}\equiv\mathbf{E}\cdot\pare{\mathbf{R}^{-1}}$, $\mathbf{E}$ is a diagonal matrix with entries $E_j$ $\pare{=k_j+i\kappa_j}$, and the matrix $\mathbf{R}$ has elements $R_{ij}$. \eqnref{eq:heff_nh} shows that the non-Hermitian nature of the effective Hamiltonian introduces cross-terms indicating polariton-polariton mixing. 

Although the single-excitation polariton operators already incorporate the nonlinear response through the $\chi^{(3)}$ dependence coming from the microscopic parameter mapping, one can explicitly see how this nonlinearity appears in a field-only representation. To this end, we adiabatically eliminate the atomic degrees of freedom by solving the Heisenberg-Langevin equations for the atomic operators, under the assumption that the material relaxes much faster than the field. We take the Langevin noise operators to be $\delta-$correlated, $\langle\hat{F}_k(t)\hat{F}_k'(t')\rangle\sim\delta\pare{t-t'}$ and drop them in the final expressions since their expectation values vanish; we further work in the weak-field regime and truncate the resulting series expansion for each atomic operator at third order in the field. A detailed description is provided in App.~\ref{app:HL_1}. Substituting these adiabatically eliminated expressions for the atomic operators into the single-excitation polariton operator yields a nonlinear field-only form $\hat{\tilde{P}}^\dag_j$, given by
\be\label{eq:nonlinearPol_1}
    \hat{\tilde{P}}^\dag_j \approx 
    \tilde{c}_{1j}^{\text{MAT}}\adop + 
    \tilde{c}_{2j}^{\text{MAT}}\adop\adop\aop,
\ee
where the coefficients $\tilde{c}_{1j}^{\text{MAT}}$ and $\tilde{c}_{2j}^{\text{MAT}}$ are expressed in terms of material susceptibilities through the same microscopic parameter mapping used to obtain the full polariton operators (Eqs.~(\ref{eq:Gamma_final})-(\ref{eq:mu_final})). However, we emphasize that this purely photonic description is strictly formal. The nonlinear photon operator cannot be identified as a genuine quasiparticle excitation of the system; rather, it simply reflects the bare field operator acquiring an effective nonlinear energy shift from the material without altering its fundamental constitution (see App.~\ref{app:HL_1}).

When the probe and signal fields are distinguishable, the microscopic model must be similarly modified such that there are two distinct electric fields $\pare{\mathcal{E}_1\neq\mathcal{E}_2}$, one coupling each atomic transition.
In this case, at least one of $\Gamma_{12}$ or $\Gamma_{21}$ must be nonzero in order to uniquely determine the zeroth-order ground-state populations.
This allows us to obtain the susceptibilities as before. The quantization process follows the single field case, with the only difference being that the single excitation subspace is now spanned by $\cpare{\adop_1\ket{1},\adop_2\ket{2},\ket{3}}$. The key difference is in the microscopic to macroscopic mapping. Now, we need to consider the polarization induced by the field $\mathcal{E}_1$ and that induced by the field $\mathcal{E}_2$ separately. In particular,
\be\label{eq:Pol_SC2_1}
\begin{split}
    \mathcal{P}_j & 
    =  \frac{N}{V}
    \mu_{3j}\rho_{3j}^{\text{ss}}
    \\
    &\approx\epsilon_0\spare{\chi^{\pare{1}}_j\mathcal{E}_j+\pare{\chi^{\pare{3}}_j}_{\text{SPM}}\abs{\mathcal{E}_j}^2\mathcal{E}_j +\pare{\chi^{\pare{3}}_j}_{\text{XPM}}\abs{\mathcal{E}_{k}}^2\mathcal{E}_j},
\end{split}
\ee
where $j,k\in\cpare{1,2}$, $k\neq j$, and the susceptibilities are obtained by Taylor expanding $\rho_{3j}^{\text{ss}}$ around both $\mathcal{E}_1=0$ and $\mathcal{E}_2=0$. We consider the set of tunable parameters within the physically allowed parameter regime to be $\cpare{\Re\pare{\chi^{(1)}_j},\Im\pare{\chi^{(1)}_j},\Re\pare{\pare{\chi^{\pare{3}}_j}_{\text{SPM}}},\Delta_1,\Delta_2}$. Within the present microscopic model, independently specifying these quantities fixes the corresponding microscopic decay rates and coupling parameters. Consequently, the remaining susceptibility components such as $\pare{\chi^{\pare{3}}_j}_{\text{XPM}}$ are determined through the following relations:
\be
    \Re\pare{\chi^{(1)}_j} = \frac{N}{V\epsilon_0}\frac{4 \Delta_j \Gamma_{kj} \mu_{3j}^2}{\hbar(\Gamma_{jk} + \Gamma_{kj}) (4 \Delta_j^2 + (\Gamma_{jk} + \Gamma_{3j} + \Gamma_{3k})^2)},
\ee
\be
    \Im\pare{\chi^{(1)}_j} = 
    \frac{\Re\pare{\chi^{(1)}_j}(\Gamma_{jk} + \Gamma_{3j} + \Gamma_{3k})}{2\Delta_j},
\ee

\begin{widetext}
\be
    \Re\pare{\pare{\chi^{\pare{3}}_j}_{\text{SPM}}} =
    -\frac{V\epsilon_0}{N}\frac{2\Re\pare{\chi^{(1)}_j}\Im\pare{\chi^{(1)}_j}(\Gamma_{jk} + 2 \Gamma_{kj} + \Gamma_{3k})}{\hbar \Gamma_{kj} (\Gamma_{3j} + \Gamma_{3k})}.
\ee
\end{widetext}

\begin{figure}
    \centering
    \includegraphics[width=\linewidth]{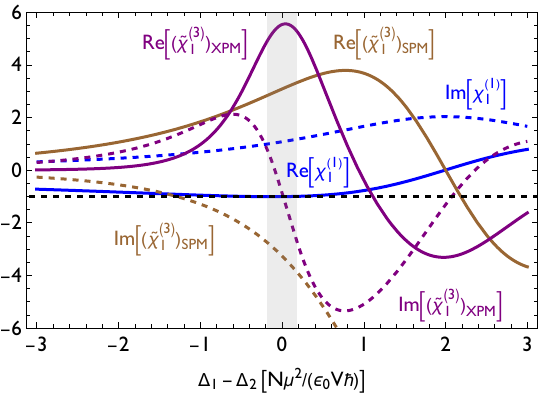}
    \caption{
    Example of the ENZ response generated by a $\Lambda$-system coupled to two fields. The shaded region highlights the relevant susceptibilities around the ENZ region. Here, $\spare{\mu}$ is defined such that $\mu_{3j}=\alpha_j\mu$, where $\alpha_j$ is a dimensionless number. The representative values of parameters chosen are $\Delta_1+\Delta_2=-2 N\mu^2/\pare{\epsilon_0V\hbar}$, $\Gamma_{31}=\Gamma_{32}=N\mu^2/\pare{\epsilon_0V\hbar}$, $\Gamma_{12}=0.12N\mu^2/\pare{\epsilon_0V\hbar}$, $\Gamma_{21} = 1.34N\mu^2/\pare{\epsilon_0V\hbar}$, $\alpha_1= 1.53$, and $\alpha_2=1.95$. $\pare{\tilde{\chi}_1^{(3)}}_{\text{SPM}}$ and $\pare{\tilde{\chi}_1^{(3)}}_{\text{XPM}}$ represent the dimensionless self- and cross-terms of the third order response with respect to the probe field.}
    \label{fig:ENZresponse_twofield}
\end{figure}

An example of the ENZ response (with respect to the probe field) generated by this model is shown in \figref{fig:ENZresponse_twofield}. With this new parameter mapping, a similar analysis as the indistinguishable field case can be done to obtain single-excitation quasiparticle operators. The key difference is that now, there are two kinds of photonic excitations, represented by $\adop_j$. A detailed analysis of this is provided in App.~\ref{app:difffields}. One can again obtain a field-only representation of the polariton operator, as in the single-field case, by adiabatically eliminating the material degrees of freedom. This yields a nonlinear single-excitation field operator of the form $\pare{\hat{\tilde{P}}^{(i)}_k}^\dag \approx \tilde{c}_{1k}^{\text{MAT}} \adop_i + \tilde{c}_{2k}^{\text{MAT}} \adop_i\adop_i\aop_i + \tilde{c}_{3k}^{\text{MAT}} \adop_i \adop_{j\neq i}\aop_{j\neq i}$,
where $k\in\cpare{1,2,3}$ and $i,j\in\cpare{1,2}$. Note that now there are two kinds of nonlinear terms, a self-term proportional to $\adop_i\adop_i\aop_i$ and a cross-term proportional to $\adop_i \adop_{j\neq i}\aop_{j\neq i}$, reflecting the fact that there are two different fields (see App.~\ref{app:HL_2} for details).

\section{Discussion and Outlook}

In this work, we set out to find a quantized description of a lossy nonlinear ENZ material. Our approach utilizes a solvable microscopic model with a steady-state response that can be matched to that of an ENZ-like medium. Within a projected non-Hermitian formulation, we obtain polariton operators whose coefficients are determined by macroscopic susceptibilities. By adiabatically eliminating the material degrees of freedom, we further obtain a quantized field operator that contains a linear term proportional to $\adop$ and a nonlinear term proportional to $\adop\adop\aop$, the relative weights of which are given by $\tilde{c}_{1j}^{\text{MAT}}$ and $\tilde{c}_{2j}^{\text{MAT}}$ which depend on material susceptibilities $\chi^{(1)}$ and $\chi^{(3)}$ (see \eqnref{eq:nonlinearPol_1}). This field-only representation is the closest analogue, within our framework, to the conventional descriptions of Kerr media, where the nonlinear response is encoded through an effective interaction proportional to $\adop\adop\aop\aop$ ~\cite{Hillery_2009,DrummondHillery}. However, the ENZ case differs from the usual setting in which one begins with well-defined linear polaritons and treats the nonlinearity as interactions between them; since the real part of the linear permittivity vanishes in the ENZ regime, the existence and constitution of the elementary excitation cannot be assumed a priori. 

Our microscopic polariton-based quantization reveals that in the single-excitation regime, the elementary excitations are approximately bosonic quasiparticles in which the excitation is stored partly in the field and partly in the material. The microscopic model should not be interpreted as a literal description of the ENZ material; rather, it serves as a solvable scaffold that realizes a class of effective responses consistent with the prescribed values of $\chi^{(1)}$ and $\chi^{(3)}$. Such a description incorporates losses through the imaginary part of $\chi^{\pare{1}}$, giving the quasiparticles a finite lifetime and producing deviations from perfectly bosonic behavior. The nonlinearity, likewise, is included in the single-excitation description through the $\chi^{(3)}$ dependence coming from the microscopic parameter mapping. Thus, the full 
polariton construction answers the central question of this work: in a lossy nonlinear ENZ medium, a single quantum of excitation is a dressed polariton whose field and material content is set by $\chi^{(1)}$, $\chi^{(3)}$, and the loss.

We emphasize that the purely photonic description is strictly formal. The nonlinear photon operator is not an independent canonical quantization of the nonlinear ENZ field and should not be identified as a genuine quasiparticle excitation of the system.  Instead, it shows how, after eliminating the material degrees of freedom, the bare photon creation operator acquires an effective nonlinear term proportional to $\adop\adop\aop$, with coefficients $\tilde{c}_{1j}^{\text{MAT}}$ and $\tilde{c}_{2j}^{\text{MAT}}$ fixed by the same macroscopic susceptibilities that determine the full polariton operator. The field-only expression, therefore, connects our results to traditional nonlinear-optics language, while the full polariton operator retains the physical description of the coupled light-matter excitation.

Several limitations of this framework should be emphasized. Throughout this work, we employ a single-mode approximation; the framework should therefore be understood as a quantization of the relevant resonant mode, rather than a complete continuum quantization of all relevant electromagnetic modes in the medium. In addition, a fully canonical quantization of a lossy nonlinear ENZ material has not been achieved here; our quantization scheme is an approximation, strictly valid only when a non-Hermitian Hamiltonian formulation is appropriate (i.e., re-population of the ground states does not qualitatively alter the dynamics). This approximation is however valid in the low-excitation limit considered in this work. 

Despite these limitations, our approach provides useful insight into how the polariton operator is modified in the presence of loss and nonlinearity. It also establishes a representative quantization procedure, consistent with measured susceptibility values, without requiring an actual microscopic description of the ENZ material. Natural extensions of this work include going beyond the single-mode approximation and incorporating spatial inhomogeneity and anisotropy. Such generalizations will be important for developing a more complete quantum description of realistic ENZ materials. Another interesting direction would involve switching from the discrete-level system used here to a continuous band structure model inherent to real ENZ media in order to obtain a fully quantized version of $\chi^{(3)}$ from first principles.

\textbf{Acknowledgments} We thank Muhammad Miskeen Khan, Michael Fleischhauer, and Alex Gong for useful discussions. S.F.Y., D.G., and S.K.O. acknowledge the QEMISD MURI project funded by ARO-W911NF-24-2-0195 P00001. S.F.Y. also acknowledges NSF via
the CUA PFC PHY-2317134, and QuSeC-TAQS OMA2326787 in addition to AFOSR FA9550-24-1-0311. A.P. ackowledges the Harvard Purcell Fellowship. D.G. and S.K.O. also acknowledge the support by ARO W911NF 2510283.

\onecolumngrid
\appendix

\section{Exact expressions for the relative weights of the polariton constituents}\label{app:PolaritonCoeffs}

The relative weights of the polariton constituents $c_{ij}$, in \eqnref{eq:Adag} and \eqnref{eq:Bdag}, are given by
\be\begin{split}
    c_{1j} &= \frac{1}{\sqrt{1+\frac{\pare{k_j^2+\kappa_j^2}\pare{g_2^2+\pare{k_j+\Delta_1-\Delta_2}^2+\kappa_j^2}}{g_1^2\pare{\pare{k_j+\Delta_1-\Delta_2}^2+\kappa_j^2}}}},\\
    c_{2j} &= \frac{g_2\pare{k_j+i\kappa_j}}{g_1\pare{k_j+\Delta_1-\Delta_2+i\kappa_j}\sqrt{1+\frac{\pare{k_j^2+\kappa_j^2}\pare{g_2^2+\pare{k_j+\Delta_1-\Delta_2}^2+\kappa_j^2}}{g_1^2\pare{\pare{k_j+\Delta_1-\Delta_2}^2+\kappa_j^2}}}},\\
    c_{3j} &= \frac{k_j+i\kappa_j}{g_1\sqrt{1+\frac{\pare{k_j^2+\kappa_j^2}\pare{g_2^2+\pare{k_j+\Delta_1-\Delta_2}^2+\kappa_j^2}}{g_1^2\pare{\pare{k_j+\Delta_1-\Delta_2}^2+\kappa_j^2}}}},
\end{split}\ee
where $k_j\equiv\Re\pare{E_j}$ and $\kappa_j\equiv\Im\pare{E_j}$. The overall phase of the coefficients $c_{ij}$ is chosen such that $c_{1j} \in \mathbb{R}$ and positive.

\section{Adiabatic elimination of the atomic operators to obtain a nonlinear photon operator}

\subsection{Indistinguishable Signal and Probe Fields}\label{app:HL_1}

The time evolution for each  atomic operator $\hat{\sigma}_{ij}(t)$, given by the Heisenberg-Langevin equation, can be written as
\begin{equation*}
\begin{split}
    \frac{d\hat{\sigma}_{11}}{dt} &=i\pare{g_1 ^*\hat{a}^\dag\hat{\sigma}_{13} - g_1\hat{\sigma}_{31} \hat{a}}+
    \Gamma_{31} \hat{\sigma}_{33}+ \sqrt{\Gamma_{31}}\hat{F}_{1}(t),
\end{split}
\end{equation*}

\begin{equation*}
\begin{split}
    \frac{d\hat{\sigma}_{12}}{dt} &=-i\pare{\pare{\Delta_1-\Delta_2}\hat{\sigma}_{12}-g_2^*\hat{a}^\dag \hat{\sigma}_{13}+g_1 \hat{\sigma}_{32}\hat{a}},
\end{split}
\end{equation*}

\begin{equation*}
\begin{split}
    \frac{d\hat{\sigma}_{13}}{dt} &=-i\pare{\Delta_1 \hat{\sigma}_{13}
    -g_1 \hat{\sigma}_{11}\hat{a}-g_2 \hat{\sigma}_{12} \hat{a}+ g_1 \hat{\sigma}_{33} \hat{a}}
    -\frac{\Gamma_{31}+\Gamma_{32}}{2}\hat{\sigma}_{13} + \sqrt{\frac{\Gamma_{31}+\Gamma_{32}}{2}}\hat{F}_{13}(t),
\end{split}
\end{equation*}

\begin{equation*}
\begin{split}
    \frac{d\hat{\sigma}_{21}}{dt} &=i\pare{
    \pare{\Delta_1-\Delta_2}\hat{\sigma}_{21}+g_1 ^*\hat{a}^\dag\hat{\sigma}_{23}-g_2\hat{\sigma}_{31}\hat{a}},
\end{split}
\end{equation*}

\begin{equation*}
\begin{split}
    \frac{d\hat{\sigma}_{22}}{dt} &= i\pare{g_2^*\hat{a}^\dag \hat{\sigma}_{23} - g_2 \hat{\sigma}_{32}\hat{a}}
    +\Gamma_{32} \hat{\sigma}_{33}+ \sqrt{\Gamma_{32}}\hat{F}_2(t),
\end{split}
\end{equation*}

\begin{equation*}
\begin{split}
    \frac{d\hat{\sigma}_{23}}{dt} &= -i \pare{\Delta_2\hat{\sigma}_{23}
    - g_1 \hat{\sigma}_{21} \hat{a}
    - g_2 \hat{\sigma}_{22} \hat{a}
    + g_2 \hat{\sigma}_{33} \hat{a}}
    -\frac{\Gamma_{31}+\Gamma_{32}}{2} \hat{\sigma}_{23}+ \sqrt{\frac{\Gamma_{31}+\Gamma_{32}}{2}}\hat{F}_{23}(t),
\end{split}
\end{equation*}

\begin{equation*}
\begin{split}
    \frac{d\hat{\sigma}_{31}}{dt} &=i
    \pare{\Delta_1\hat{\sigma}_{31} 
    + g_1 ^*\hat{a}^\dag\hat{\sigma}_{33} 
    - g_1 ^*\hat{a}^\dag\hat{\sigma}_{11} 
    - g_2^*\hat{a}^\dag \hat{\sigma}_{21}}
    - \frac{\Gamma_{31}+\Gamma_{32}}{2}\hat{\sigma}_{31}
    + \sqrt{\frac{\Gamma_{31}+\Gamma_{32}}{2}}\hat{F}_{31}(t),
\end{split}
\end{equation*}

\begin{equation*}
\begin{split}
    \frac{d\hat{\sigma}_{32}}{dt} &=i
    \pare{\Delta_2\hat{\sigma}_{32}+g_2^*\hat{a}^\dag \hat{\sigma}_{33} - g_1^* \hat{a}^\dag\hat{\sigma}_{12} - g_2^*\hat{a}^\dag\hat{\sigma}_{22}}-\frac{\Gamma_{31}+\Gamma_{32}}{2}\hat{\sigma}_{32} + \sqrt{\frac{\Gamma_{31}+\Gamma_{32}}{2}}\hat{F}_{32}(t),
\end{split}
\end{equation*}

\begin{equation*}
\begin{split}
    \frac{d\hat{\sigma}_{33}}{dt} &=i\pare{
      g_1 \hat{\sigma}_{31}\hat{a}
    + g_2 \hat{\sigma}_{32}\hat{a}
    - g_1 ^*\hat{a}^\dag\hat{\sigma}_{13} 
    - g_2^*\hat{a}^\dag \hat{\sigma}_{23}
    } -\pare{\Gamma_{31}+\Gamma_{32}}\hat{\sigma}_{33} + \sqrt{\Gamma_{31}+\Gamma_{32}}\hat{F}_{33}(t).
\end{split}
\end{equation*}
We can now solve the equations perturbatively up to third order in $\hat{a}^{\dag}$, with $\hat{\sigma}_{ij} \equiv \sum\limits_{n=0}^\infty \hat{\sigma}_{ij}^{(n)} = \sum\limits_{n=0}^\infty\sum\limits_{k\leq n} \sigma_{ij}^{(n)}\pare{\hat{a}^\dag}^k\hat{a}^{n-k}$, under the adiabatic approximation.  We also assume that the noise operators are delta-correlated noise operators $\langle\hat{F}_k(t)\hat{F}_k'(t')\rangle\sim\delta\pare{t-t'}$ and drop them in this perturbative adiabatic treatment. Although the atomic and photonic operators act on different parts of the full Hilbert space prior to elimination, the adiabatically eliminated atomic operators should be understood as their steady-state values determined by the slow photonic dynamics. The resulting relations between the atomic and photonic operators are therefore effective relations within the reduced slow manifold, not exact operator identities on the full Hilbert space.

To zeroth order, only $\sigma_{11}, \sigma_{22}\neq 0$, and $\sigma_{11}+\sigma_{22}=1$ ($\sigma_{11}\equiv\langle\hat{\sigma}^{(0)}_{11}\rangle$ and $\sigma_{22}\equiv\langle\hat{\sigma}^{(0)}_{22}\rangle$, without the hats, are numbers). In the adiabatic limit, we can replace $\hat{\sigma}_{11}^{(0)}\to\sigma_{11}$ and $\hat{\sigma}_{22}^{(0)}\to\sigma_{22}$. Here, the zeroth-order populations are not arbitrary populations in the degenerate ground-state manifold, but should be understood in the same finite-drive limiting sense $\mathcal{E}\to 0$. The weak field lifts this degeneracy through optical pumping, resulting in
\begin{equation*}
\begin{split}
   \sigma_{11}^{(0)} = \frac{\Gamma_{31}\abs{g_2}^2\pare{\Delta_1^2+\pare{\frac{\Gamma_{31}+\Gamma_{32}}{2}}^2}}{\Gamma_{32}\abs{g_1}^2\pare{\Delta_2^2+\pare{\frac{\Gamma_{31}+\Gamma_{32}}{2}}^2}+\Gamma_{31}\abs{g_2}^2\pare{\Delta_1^2+\pare{\frac{\Gamma_{31}+\Gamma_{32}}{2}}^2}} \equiv \sigma_{11},
\end{split}
\end{equation*}
\begin{equation*}
\begin{split}
   \sigma_{22}^{(0)} = \frac{\Gamma_{32}\abs{g_1}^2\pare{\Delta_2^2+\pare{\frac{\Gamma_{31}+\Gamma_{32}}{2}}^2}}{\Gamma_{32}\abs{g_1}^2\pare{\Delta_2^2+\pare{\frac{\Gamma_{31}+\Gamma_{32}}{2}}^2}+\Gamma_{31}\abs{g_2}^2\pare{\Delta_1^2+\pare{\frac{\Gamma_{31}+\Gamma_{32}}{2}}^2}} \equiv \sigma_{22}.
\end{split}
\end{equation*}

The non-zero terms at first order are given by:
\begin{equation*}
\begin{split}
    \hat{\sigma}_{13}^{(1)} = \frac{2ig_1 \sigma_{11}}{\Gamma_{31}+\Gamma_{32}+2i\Delta_1}\hat{a},
\end{split}
\end{equation*}

\begin{equation*}
\begin{split}
    \hat{\sigma}_{23}^{(1)} = \frac{2ig_2 \pare{1-\sigma_{11}}}{\Gamma_{31}+\Gamma_{32}+2i\Delta_2}\hat{a},
\end{split}
\end{equation*}

\begin{equation*}
\begin{split}
    \hat{\sigma}_{31}^{(1)} = -\frac{2ig_1^* \sigma_{11}}{\Gamma_{31}+\Gamma_{32}-2i\Delta_1}\hat{a}^\dag,
\end{split}
\end{equation*}

\begin{equation*}
\begin{split}
    \hat{\sigma}_{32}^{(1)} = -\frac{2ig_2^* \pare{1-\sigma_{11}}}{\Gamma_{31}+\Gamma_{32}-2i\Delta_2}\hat{a}^\dag.
\end{split}
\end{equation*}

The non-zero terms at second order are given by:
\begin{equation*}
\begin{split}
    \hat{\sigma}_{12}^{(2)} = \frac{2ig_1g_2^*\pare{\abs{g_1}^2 \Gamma_{32} (\Gamma_{31} + \Gamma_{32} + 
    2i\Delta_2) + \abs{g_2}^2 \Gamma_{31} (\Gamma_{31} + \Gamma_{32} - 
    2i\Delta_1)}\hat{a}^\dag \hat{a}}{\pare{\Delta_1-\Delta_2}\pare{\abs{g_1}^2 \Gamma_{32} ((\Gamma_{31} + \Gamma_{32})^2 + 4 \Delta_2^2) + \abs{g_2}^2 \Gamma_{31} ((\Gamma_{31} + \Gamma_{32})^2 + 4 \Delta_1^2)}},
\end{split}
\end{equation*}

\begin{equation*}
\begin{split}
    \hat{\sigma}_{21}^{(2)} = -\frac{2ig_1^*g_2\pare{\abs{g_1}^2 \Gamma_{32} (\Gamma_{31} + \Gamma_{32} - 2i\Delta_2) + \abs{g_2}^2 \Gamma_{31} (\Gamma_{31} + \Gamma_{32} + 2i\Delta_1) }\hat{a}^\dag \hat{a}}{\pare{\Delta_1-\Delta_2}\pare{\abs{g_1}^2 \Gamma_{32} ((\Gamma_{31} + \Gamma_{32})^2 + 4 \Delta_2^2) + \abs{g_2}^2 \Gamma_{31} ((\Gamma_{31} + \Gamma_{32})^2 + 4 \Delta_1^2)}},
\end{split}
\end{equation*}

\begin{equation*}
\begin{split}
    \hat{\sigma}_{33}^{(2)} = \frac{4\abs{g_1}^2\abs{g_2}^2\pare{\Gamma_{31}+\Gamma_{32}}}{\abs{g_1}^2\Gamma_{32}\pare{\pare{\Gamma_{31}+\Gamma_{32}}^2+4\Delta_2^2}+\abs{g_2}^2\Gamma_{31}\pare{\pare{\Gamma_{31}+\Gamma_{32}}^2+4\Delta_1^2}}\hat{a}^\dag\hat{a},
\end{split}
\end{equation*}

\begin{equation}\begin{split}\label{eq:sigma11_2}
    &\hat{\sigma}_{11} ^{(2)} = \frac{\Sigma_{11}^{(2)}\hat{a}^\dag \hat{a}}{(\Delta_1 - \Delta_2) (\abs{g_1}^2 \Gamma_{32} ((\Gamma_{31} + \Gamma_{32})^2 + 4 \Delta_2^2) + \abs{g_2}^2 \Gamma_{31} ((\Gamma_{31} + \Gamma_{32})^2 + 4 \Delta_1^2))^2},
\end{split}\end{equation}
where $\Sigma_{11}^{(2)} = -4 \abs{g_2}^2 \Gamma_{31} (\Gamma_{32} (-(\Gamma_{31} + \Gamma_{32})^2 \Delta_1 + (3 (\Gamma_{31} + \Gamma_{32})^2 + 8 \Delta_1^2) \Delta_2 - 4 \Delta_1 \Delta_2^2 + 4 \Delta_2^3) \abs{g_1}^4+\abs{g_1}^2 \abs{g_2}^2 ((\Gamma_{31} + \Gamma_{32})^2 (\Gamma_{31} + 4 \Gamma_{32}) \Delta_1 + 4 (\Gamma_{31} + 2 \Gamma_{32}) \Delta_1^3 - (\Gamma_{31} + 2 \Gamma_{32}) ((\Gamma_{31} + \Gamma_{32})^2 + 4 \Delta_1^2) \Delta_2 + 8 \Gamma_{32} \Delta_1 \Delta_2^2))$.
\begin{equation}\begin{split}\label{eq:sigma22_2}
    &\hat{\sigma}_{22} ^{(2)} = \frac{\Sigma_{22}^{(2)}\hat{a}^\dag \hat{a}}{(\Delta_1 - \Delta_2) (\abs{g_1}^2 \Gamma_{32} ((\Gamma_{31} + \Gamma_{32})^2 + 4 \Delta_2^2) + \abs{g_2}^2\Gamma_{31} ((\Gamma_{31} + \Gamma_{32})^2 + 4 \Delta_1^2))^2},
\end{split}\end{equation}
where $\Sigma_{22}^{(2)} = 4 \Gamma_{32} (\abs{g_1}^2 \abs{g_2}^4 \Gamma_{31} (3 (\Gamma_{31} + \Gamma_{32})^2 \Delta_1 + 4 \Delta_1^3 - ((\Gamma_{31} + \Gamma_{32})^2 + 4 \Delta_1^2) \Delta_2 + 8 \Delta_1 \Delta_2^2)+ \abs{g_1}^4 \abs{g_2}^2(-(\Gamma_{31} + \Gamma_{32})^2 (2 \Gamma_{31} + \Gamma_{32}) \Delta_1 + (\Gamma_{31} + \Gamma_{32})^2 (4 \Gamma_{31} + \Gamma_{32}) \Delta_2 + 8 \Gamma_{31} \Delta_1^2 \Delta_2 - 4 (2 \Gamma_{31} + \Gamma_{32}) \Delta_1 \Delta_2^2 + 4 (2 \Gamma_{31} + \Gamma_{32}) \Delta_2^3))$.

The non-zero terms at third order are given by:
\be\begin{split}
    \hat{\sigma}_{13} ^{(3)} &= \frac{\Sigma_{13}^{(3)}\hat{a}^\dag\hat{a}\hat{a}}{(\Delta_1 - \Delta_2) (\abs{g_1}^2 \Gamma_{32} ((\Gamma_{31} + \Gamma_{32})^2 + 4 \Delta_2^2) + \abs{g_2}^2 \Gamma_{31} ((\Gamma_{31} + \Gamma_{32})^2 + 4 \Delta_1^2))^2},
\end{split}\ee
where $\Sigma_{13}^{(3)} = -4 g_1 (\Gamma_{31}^2 (\Gamma_{31} + \Gamma_{32} - 2 i \Delta_1)^2 \abs{g_2}^6 +\abs{g_1}^4 \abs{g_2}^2\Gamma_{32} (\Gamma_{32} (\Gamma_{31} + \Gamma_{32})^2 + 4 i \Gamma_{31} (\Gamma_{31} + \Gamma_{32} - 2 i \Delta_1) \Delta_2 + 4 \Gamma_{32} \Delta_2^2)+2 \abs{g_1}^2 \abs{g_2}^4\Gamma_{31} (\Gamma_{31}^2 \Gamma_{32} + 2 \Gamma_{31} \Gamma_{32}^2 + \Gamma_{32}^3 + 2 i \Gamma_{31}^2 \Delta_1 + 4 i \Gamma_{31} \Gamma_{32} \Delta_1 + 2 i \Gamma_{32}^2 \Delta_1 + 4 \Gamma_{31} \Delta_1^2 + 6 \Gamma_{32} \Delta_1^2 - 2 i (\Gamma_{31} + \Gamma_{32}) (\Gamma_{31} + \Gamma_{32} - 2 i \Delta_1) \Delta_2 + 2 \Gamma_{32} \Delta_2^2))$.
\be\begin{split}
    \hat{\sigma}_{23} ^{(3)} &= \frac{\Sigma_{23}^{(3)}\hat{a}^\dag\hat{a}\hat{a}}{(\Delta_1 - \Delta_2) (\abs{g_1}^2 \Gamma_{32} ((\Gamma_{31} + \Gamma_{32})^2 + 4 \Delta_2^2) + \abs{g_2}^2 \Gamma_{31} ((\Gamma_{31} + \Gamma_{32})^2 + 4 \Delta_1^2))^2},\\
\end{split}\ee
where $\Sigma_{23}^{(3)}=4 g_2 (\Gamma_{32}^2 (\Gamma_{31} + \Gamma_{32} - 2 i \Delta_2)^2 \abs{g_1}^6 + \abs{g_1}^2 \abs{g_2}^4\Gamma_{31} (\Gamma_{31}^3 + 2 \Gamma_{31}^2 \Gamma_{32} + \Gamma_{31} (\Gamma_{32}^2 + 4 i \Gamma_{32} \Delta_1 + 4 \Delta_1^2) + 4 \Gamma_{32} \Delta_1 (i \Gamma_{32} + 2 \Delta_2)) +2 \abs{g_1}^4 \abs{g_2}^2\Gamma_{32} (\Gamma_{31}^3 + 2 \Gamma_{31}^2 (\Gamma_{32} - i \Delta_1 + i \Delta_2) - 2 \Gamma_{32} (\Delta_1 - \Delta_2) (i \Gamma_{32} + 2 \Delta_2) + \Gamma_{31} (\Gamma_{32}^2 -  4 i \Gamma_{32} \Delta_1 + 2 \Delta_1^2 + 4 i \Gamma_{32} \Delta_2 - 4 \Delta_1 \Delta_2 + 6 \Delta_2^2)))$.
\be
    \hat{\sigma}_{31} ^{(3)} = \frac{\Sigma_{31}^{(3)}\hat{a}^\dag\hat{a}^\dag\hat{a}}{(\Delta_1 - \Delta_2) (\abs{g_1}^2 \Gamma_{32} ((\Gamma_{31} + \Gamma_{32})^2 + 4 \Delta_2^2) + \abs{g_2}^2 \Gamma_{31} ((\Gamma_{31} + \Gamma_{32})^2 + 4 \Delta_1^2))^2},
\ee
where $\Sigma_{31}^{(3)}=-4 g_1^* (\Gamma_{31}^2 (\Gamma_{31} + \Gamma_{32} + 2 i \Delta_1)^2 \abs{g_2}^6 + 2 \abs{g_1}^2 \abs{g_2}^4 \Gamma_{31} (\Gamma_{31}^2 \Gamma_{32} + 2 \Gamma_{31} \Gamma_{32}^2 + \Gamma_{32}^3 - 2 i \Gamma_{31}^2 \Delta_1 - 4 i \Gamma_{31} \Gamma_{32} \Delta_1 - 2 i \Gamma_{32}^2 \Delta_1 + 4 \Gamma_{31} \Delta_1^2 + 6 \Gamma_{32} \Delta_1^2 + 2 i (\Gamma_{31} + \Gamma_{32}) (\Gamma_{31} + \Gamma_{32} + 2 i \Delta_1) \Delta_2 + 2 \Gamma_{32} \Delta_2^2) + \abs{g_1}^4 \abs{g_2}^2\Gamma_{32} (\Gamma_{32} (\Gamma_{31} + \Gamma_{32})^2 - 4 i \Gamma_{31} (\Gamma_{31} + \Gamma_{32} + 2 i \Delta_1) \Delta_2 + 4 \Gamma_{32} \Delta_2^2))$.
\be
    \hat{\sigma}_{32} ^{(3)} = \frac{\Sigma_{32} ^{(3)} \hat{a}^\dag\hat{a}^\dag\hat{a}}{(\Delta_1 - \Delta_2) (\abs{g_1}^2 \Gamma_{32} ((\Gamma_{31} + \Gamma_{32})^2 + 4 \Delta_2^2) + \abs{g_2}^2 \Gamma_{31} ((\Gamma_{31} + \Gamma_{32})^2 + 4 \Delta_1^2))^2},
\ee
where $\Sigma_{32} ^{(3)} = 4 g_2^* (\Gamma_{32}^2 (\Gamma_{31} + \Gamma_{32} + 2 i \Delta_2)^2 \abs{g_1}^6 + \abs{g_1}^2  \abs{g_2}^4 \Gamma_{31} (\Gamma_{31}^3 + 2 \Gamma_{31}^2 \Gamma_{32} + \Gamma_{31} (\Gamma_{32}^2 - 4 i \Gamma_{32} \Delta_1 + 4 \Delta_1^2) + 4 \Gamma_{32} \Delta_1 (-i \Gamma_{32} + 2 \Delta_2)) + 2 \abs{g_1}^4 \abs{g_2}^2 \Gamma_{32} (\Gamma_{31}^3 + 2 \Gamma_{31}^2 (\Gamma_{32} + i (\Delta_1 - \Delta_2)) + 2 i \Gamma_{32} (\Delta_1 - \Delta_2) (\Gamma_{32} + 2 i \Delta_2) + \Gamma_{31} (\Gamma_{32}^2 + 4 i \Gamma_{32} (\Delta_1 - \Delta_2) + 2 (\Delta_1^2 - 2 \Delta_1 \Delta_2 + 3 \Delta_2^2))))$.

\begin{figure*}
     \centering
    \includegraphics[width=\textwidth]{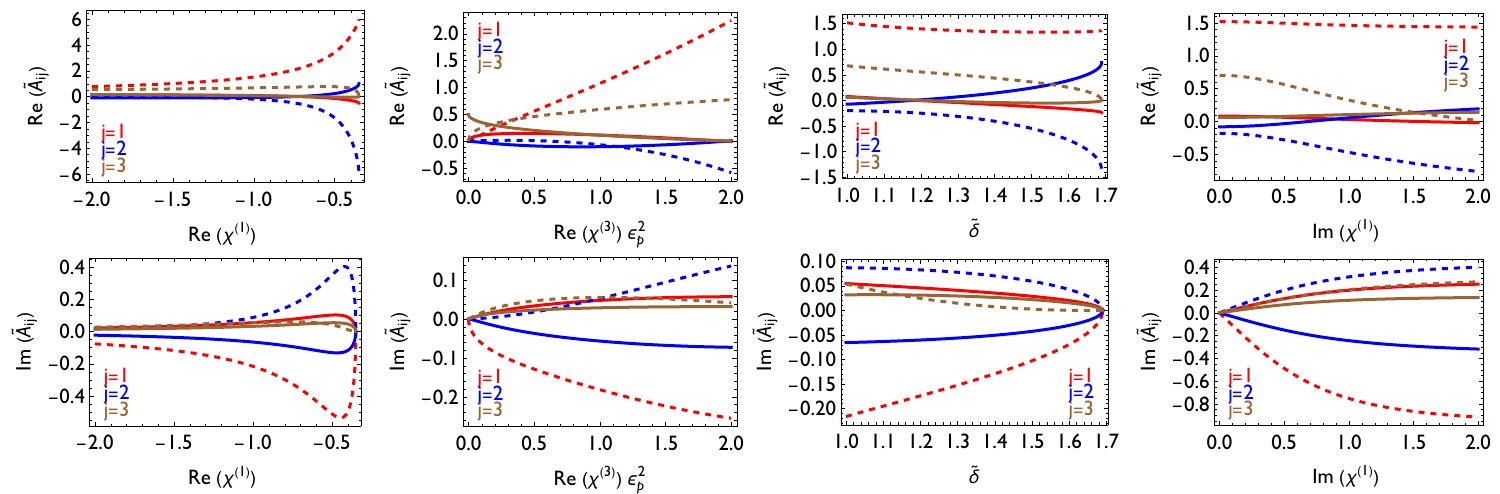}\hfill\includegraphics[width=\textwidth]{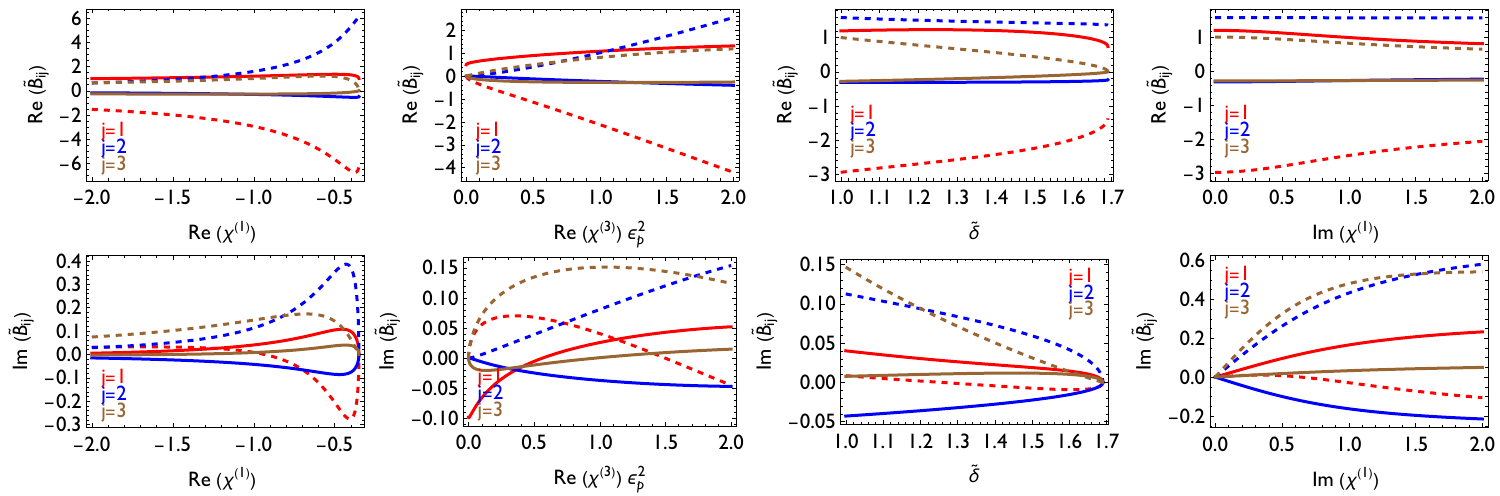}
    \caption{Real and imaginary parts of the coefficients of the field-only operator described in \eqnref{eq:PolA} and \eqnref{eq:PolB}. Solid curves show the linear coefficients, $\tilde{A}_{1j}$, $\tilde{B}_{1j}$; dashed curves show the nonlinear coefficients,  $\tilde{A}_{2j}$, $\tilde{B}_{2j}$. The representative values of parameters chosen are $\text{Re}\pare{\chi^{(1)}}=-1$, $\text{Re}\pare{\chi^{(3)}}=1.4/\epsilon_p^2$, $\tilde{\delta}=1$, and $\text{Im}\pare{\chi^{(1)}}=0.2$. In each subplot, one of these values is varied, keeping the other three fixed.}
    \label{fig:nonlinearphotonop}
\end{figure*}

If we assume, for simplicity, $\Gamma_{31}=\Gamma_{32}\equiv\Gamma$, as we did in the semiclassical case, the coefficients $\sigma_{ij}^{(n)}$ found in this quantized treatment are equivalent to the semiclassical steady-state solution $\rho_{ji}^{\pare{n}}$ with $g_j\leftrightarrow \mu_{3j}/\hbar$.
Substituting the Heisenberg-Langevin solutions for $\hat{\sigma}_{ij}$ into \eqnref{eq:Adag} and \eqnref{eq:Bdag}, we find nonlinear photon operators of the form
\be\label{eq:PolA}
    \hat{\tilde{A}}_j^\dag \approx \tilde{A}_{1j}\pare{\Delta_1,\Delta_2,\Gamma,\mu}\adop + \tilde{A}_{2j}\pare{\Delta_1,\Delta_2,\Gamma,\mu}\adop\adop\aop,
\ee
\be\label{eq:PolB}
    \hat{\tilde{B}}_j^\dag\approx \tilde{B}_{1j}\pare{\Delta_1,\Delta_2,\Gamma,\mu}\hat{a}^\dag + \tilde{B}_{2j}\pare{\Delta_1,\Delta_2,\Gamma,\mu}\adop\adop\aop,
\ee
where $\tilde{A}_{1j}\equiv c_{1j}\sigma_{11}^{(0)}+c_{3j} \sigma_{31}^{(1)}$, $\tilde{A}_{2j}\equiv c_{1j}\sigma_{11}^{(2)}+ c_{2j}\sigma_{21}^{(2)}+c_{3j} \sigma_{31}^{(3)}$, $\tilde{B}_{1j}\equiv c_{2j}\sigma_{22}^{(0)}+c_{3j} \sigma_{32}^{(1)}$, and $\tilde{B}_{2j}\equiv c_{1j}\sigma_{12}^{(2)}+ c_{2j}\sigma_{22}^{(2)} + c_{3j} \sigma_{32}^{(3)}$. Depending on the reference state chosen, $\hat{\tilde{P}}_j$, defined in \eqnref{eq:nonlinearPol_1} can be obtained from either \eqnref{eq:PolA} or \eqnref{eq:PolB}. Now, one can again use the microscopic parameter mapping to obtain $\pare{\Delta_1,\Delta_2,\Gamma,\mu}$ in terms of $\pare{\Re\pare{\chi^{(1)}},\Im\pare{\chi^{(1)}},\Re\pare{\chi^{(3)}},\delta}$, which results in \eqnref{eq:nonlinearPol_1}, with $\tilde{c}_{ij}^{\text{MAT}}\in\cpare{\tilde{A}_{ij},\tilde{B}_{ij}}$. An example of the dependencies of these coefficients on $\text{Re}\pare{\chi^{(1)}}$, $\text{Re}\pare{\chi^{(3)}}$, $\tilde{\delta}$, and $\text{Im}\pare{\chi^{(1)}}$ is shown in \figref{fig:nonlinearphotonop}.

In the coupled system, the Heisenberg equations of motion for $\aop$ and $\adop$ are given by
\be\begin{split}
    \frac{d\aop}{dt} &= \frac{i}{\hbar}\com{\hat{H}}{\aop} = i\pare{g\sigop_{13} + g\sigop_{23}}, \\
    \frac{d\adop}{dt} &= \frac{i}{\hbar}\com{\hat{H}}{\adop} = -i\pare{g\sigop_{31} + g\sigop_{32}}.
\end{split}\ee
Adiabatically eliminating the atomic degrees of freedom results in a purely photonic description of the dynamics. In particular, the equations of motion for $\aop$ and $\adop$ can be described as follows:
\be\begin{split}
    \frac{d\aop}{dt} &\approx ig\spare{\pare{\sigma_{13}^{(1)}+\sigma_{23}^{(1)}}\aop + \pare{\sigma_{13}^{(3)}+\sigma_{23}^{(3)}}\adop\aop\aop}, \\
    \frac{d\adop}{dt} &\approx -ig\spare{\pare{\sigma_{31}^{(1)}+\sigma_{32}^{(1)}}\adop + \pare{\sigma_{31}^{(3)}+\sigma_{32}^{(3)}}\adop\adop\aop}.
\end{split}\ee
The effective Hamiltonian obtained after adiabatic elimination that yields (up to third order) these equations of motion is given by
\be\begin{split}\label{eq:Heff_adel}
    \frac{H_{\text{eff}}}{\hbar} \approx &-\frac{V\epsilon_0 g^2\hbar}{N\mu^2}\spare{\pare{\Re\pare{\chi^{\pare{1}}}-i\Im\pare{\chi^{\pare{1}}}} \adop\aop + \frac{g^2\hbar^2}{2\mu^2}\pare{\Re\pare{\chi^{\pare{3}}}-i\Im\pare{\chi^{\pare{3}}}} \adop\adop\aop\aop},
\end{split}\ee
where $\chi^{\pare{1}}$ and $\chi^{\pare{3}}$ have been defined in \eqnref{eq:Pol_SC}.
Note that since there was no explicit field decay in the original system, $\aop$ and $\adop$ simply evolved under the Heisenberg equation of motion. Thus, even though the effective photon Hamiltonian is non-Hermitian, we obtain the coefficients by treating the evolution also under just the unitary Heisenberg equation of motion.

However, as noted in the main text, this formalism is not self-consistent, as it destroys the previously established polariton picture. In particular, from \eqnref{eq:Heff_adel}, it is evident that the bare photon operator $\aop^{\pare{\dag}}$ remains an eigenstate. The nonlinearity simply results in a correction to the energy $\propto \adop\aop$. 

\subsection{Distinguishable Signal and Probe Fields}\label{app:HL_2}

The time evolution for each  atomic operator $\hat{\sigma}_{ij}(t)$, given by the Heisenberg-Langevin equation, can be written as
\begin{equation*}
\begin{split}
    \frac{d\hat{\sigma}_{11}}{dt} &=i\pare{g_1 ^*\aop_1^\dag\hat{\sigma}_{13} 
    - g_1\hat{\sigma}_{31} \aop_1}
    + \Gamma_{31} \hat{\sigma}_{33}
    -\Gamma_{12}\hat{\sigma}_{11}
    +\Gamma_{21}\hat{\sigma}_{22}
    + \sqrt{\Gamma_{31}}\hat{F}_{1}(t),
\end{split}
\end{equation*}

\begin{equation*}
\begin{split}
    \frac{d\hat{\sigma}_{12}}{dt} &=-i\pare{\pare{\Delta_1-\Delta_2}\hat{\sigma}_{12}-g_2^*\aop_2^\dag \hat{\sigma}_{13}+g_1 \hat{\sigma}_{32}\aop_1}
    -\frac{\Gamma_{12}+\Gamma_{21}}{2}\hat{\sigma}_{12}
    + \sqrt{\frac{\Gamma_{12}+\Gamma_{21}}{2}}\hat{F}_{12}(t),
\end{split}
\end{equation*}

\begin{equation*}
\begin{split}
    \frac{d\hat{\sigma}_{13}}{dt} &=-i\pare{\Delta_1 \hat{\sigma}_{13}
    -g_1 \hat{\sigma}_{11}\aop_1-g_2 \hat{\sigma}_{12} \aop_2+ g_1 \hat{\sigma}_{33} \aop_1}
    -\frac{\Gamma_{31}+\Gamma_{32}+\Gamma_{12}}{2}\hat{\sigma}_{13} 
    + \sqrt{\frac{\Gamma_{31}+\Gamma_{32}+\Gamma_{12}}{2}}\hat{F}_{13}(t),
\end{split}
\end{equation*}

\begin{equation*}
\begin{split}
    \frac{d\hat{\sigma}_{21}}{dt} &=i\pare{
    \pare{\Delta_1-\Delta_2}\hat{\sigma}_{21}+g_1 ^*\aop_1^\dag\hat{\sigma}_{23}-g_2\hat{\sigma}_{31}\aop_2}
    -\frac{\Gamma_{12}+\Gamma_{21}}{2}\hat{\sigma}_{21}
    + \sqrt{\frac{\Gamma_{12}+\Gamma_{21}}{2}}\hat{F}_{21}(t),
\end{split}
\end{equation*}

\begin{equation*}
\begin{split}
    \frac{d\hat{\sigma}_{22}}{dt} &= i\pare{g_2^*\aop_2^\dag \hat{\sigma}_{23} 
    - g_2 \hat{\sigma}_{32}\aop_2}
    +\Gamma_{32} \hat{\sigma}_{33}
    +\Gamma_{12}\hat{\sigma}_{11}
    -\Gamma_{21}\hat{\sigma}_{22}
    + \sqrt{\Gamma_{32}}\hat{F}_2(t),
\end{split}
\end{equation*}

\begin{equation*}
\begin{split}
    \frac{d\hat{\sigma}_{23}}{dt} &= -i \pare{\Delta_2\hat{\sigma}_{23}
    - g_1 \hat{\sigma}_{21} \aop_1
    - g_2 \hat{\sigma}_{22} \aop_2
    + g_2 \hat{\sigma}_{33} \aop_2}
    -\frac{\Gamma_{31}+\Gamma_{32}+\Gamma_{21}}{2} \hat{\sigma}_{23}
    + \sqrt{\frac{\Gamma_{31}+\Gamma_{32}+\Gamma_{21}}{2}}\hat{F}_{23}(t),
\end{split}
\end{equation*}

\begin{equation*}
\begin{split}
    \frac{d\hat{\sigma}_{31}}{dt} &=i
    \pare{\Delta_1\hat{\sigma}_{31} 
    + g_1 ^*\aop_1^\dag\hat{\sigma}_{33} 
    - g_1 ^*\aop_1^\dag\hat{\sigma}_{11} 
    - g_2^*\aop_2^\dag \hat{\sigma}_{21}}
    - \frac{\Gamma_{31}+\Gamma_{32}+\Gamma_{12}}{2}\hat{\sigma}_{31}
    + \sqrt{\frac{\Gamma_{31}+\Gamma_{32}+\Gamma_{12}}{2}}\hat{F}_{31}(t),
\end{split}
\end{equation*}

\begin{equation*}
\begin{split}
    \frac{d\hat{\sigma}_{32}}{dt} &=i
    \pare{\Delta_2\hat{\sigma}_{32}+g_2^*\aop_2^\dag \hat{\sigma}_{33} - g_1^* \aop_1^\dag\hat{\sigma}_{12} - g_2^*\aop_2^\dag\hat{\sigma}_{22}}-\frac{\Gamma_{31}+\Gamma_{32}+\Gamma_{21}}{2}\hat{\sigma}_{32} 
    + \sqrt{\frac{\Gamma_{31}+\Gamma_{32}+\Gamma_{21}}{2}}\hat{F}_{32}(t),
\end{split}
\end{equation*}

\begin{equation*}
\begin{split}
    \frac{d\hat{\sigma}_{33}}{dt} &=i\pare{
      g_1 \hat{\sigma}_{31}\aop_1
    + g_2 \hat{\sigma}_{32}\aop_2
    - g_1 ^*\aop_1^\dag\hat{\sigma}_{13} 
    - g_2^*\aop_2^\dag \hat{\sigma}_{23}
    } -\pare{\Gamma_{31}+\Gamma_{32}}\hat{\sigma}_{33} 
    + \sqrt{\Gamma_{31}+\Gamma_{32}}\hat{F}_{33}(t).
\end{split}
\end{equation*}
We can now solve the equations perturbatively up to third order in the fields, with $\hat{\sigma}_{ij} \equiv \sum\limits_{n=0}^\infty \hat{\sigma}_{ij}^{(n)} \equiv \sum\limits_{n=0}^\infty \sigma_{ij}^{(n)}\pare{\hat{f}^\dag}^k\hat{f}^l$, $k+l=n$, under the adiabatic approximation, where $\hat{f}\in\cpare{\aop_1,\aop_2}$.  We also assume that the noise operators are delta-correlated noise operators $\langle\hat{F}_k(t)\hat{F}_k'(t')\rangle\sim\delta\pare{t-t'}$ and drop them in this perturbative adiabatic treatment.

The only non-zero terms at zeroth order are $\hat{\sigma}_{11}^{(0)}$ and $\hat{\sigma}_{22}^{(0)}$ and are given by:
\begin{equation*}\begin{split}
    \hat{\sigma}_{11}^{(0)}= \frac{\Gamma_{21}}{\Gamma_{12}+\Gamma_{21}}\mathbb{1},
\end{split}\end{equation*}
\begin{equation*}\begin{split}
    \hat{\sigma}_{22}^{(0)}= \frac{\Gamma_{12}}{\Gamma_{12}+\Gamma_{21}}\mathbb{1}.
\end{split}\end{equation*}

The non-zero terms at first order are given by:
\begin{equation*}
\begin{split}
    \hat{\sigma}_{13}^{(1)}
    =\frac{2g_1\Gamma_{21}}{\pare{\Gamma_{12}+\Gamma_{21}}\pare{2\Delta_1-i\pare{\Gamma_{31}+\Gamma_{32}+\Gamma_{12}}}}\hat{a}_1,
\end{split}
\end{equation*}

\begin{equation*}
\begin{split}
    \hat{\sigma}_{23}^{(1)}
    =\frac{2g_2\Gamma_{12}}{\pare{\Gamma_{12}+\Gamma_{21}}\pare{2\Delta_2-i\pare{\Gamma_{31}+\Gamma_{32}+\Gamma_{21}}}}\hat{a}_2,
\end{split}
\end{equation*}

\begin{equation*}
\begin{split}
    \hat{\sigma}_{31}^{(1)} =\frac{2g_1^* \Gamma_{21}}{\pare{\Gamma_{12}+\Gamma_{21}}\pare{2\Delta_1+i\pare{\Gamma_{31}+\Gamma_{32}+\Gamma_{12}}}}\hat{a}^\dag_1,
\end{split}
\end{equation*}

\begin{equation*}
\begin{split}
    \hat{\sigma}_{32}^{(1)} = \frac{2g_2^* \Gamma_{12}}{\pare{\Gamma_{12}+\Gamma_{21}}\pare{2\Delta_2+i\pare{\Gamma_{31}+\Gamma_{32}+\Gamma_{21}}}}\hat{a}^\dag_2.
\end{split}
\end{equation*}

The non-zero terms at second order are given by:
\begin{equation*}
\begin{split}
    \hat{\sigma}_{12}^{(2)}
    = \frac{4 g_1 g_2^* (-2 \Delta_1 \Gamma_{12} + i (\Gamma_{12}^2 + \Gamma_{12} (\Gamma_{31} + \Gamma_{32}) + \Gamma_{21} (-2 i \Delta_2 + \Gamma_{21} + \Gamma_{31} + \Gamma_{32}))) }{(\Gamma_{12} + \Gamma_{21}) (2 \Delta_1 - 2 \Delta_2 - i (\Gamma_{12} + \Gamma_{21})) (2 \Delta_1 - i (\Gamma_{12} + \Gamma_{31} + \Gamma_{32})) (2 \Delta_2 + i (\Gamma_{21} + \Gamma_{31} + \Gamma_{32}))}\adop_2\aop_1,
\end{split}
\end{equation*}

\begin{equation*}
\begin{split}
    \hat{\sigma}_{21}^{(2)} = -\frac{4 g_1^* g_2 (2 \Delta_1 \Gamma_{12} + i (\Gamma_{12}^2 + \Gamma_{12} (\Gamma_{31} + \Gamma_{32}) + \Gamma_{21} (2 i \Delta_2 + \Gamma_{21} + \Gamma_{31} + \Gamma_{32})))}{(\Gamma_{12} + \Gamma_{21}) (2 \Delta_1 - 2 \Delta_2 + i (\Gamma_{12} + \Gamma_{21})) (2 \Delta_1 + i (\Gamma_{12} + \Gamma_{31} + \Gamma_{32})) (2 \Delta_2 - i (\Gamma_{21} + \Gamma_{31} + \Gamma_{32}))}\adop_1\aop_2,
\end{split}
\end{equation*}

\begin{equation*}
\begin{split}
    \hat{\sigma}_{33}^{(2)} &= 
    \frac{4 \abs{g_1}^2 \Gamma_{21} (\Gamma_{12} + \Gamma_{31} + \Gamma_{32})}{(\Gamma_{12} + \Gamma_{21}) (\Gamma_{31} + \Gamma_{32}) (-2 i \Delta_1 + \Gamma_{12} + \Gamma_{31} + \Gamma_{32}) (2 i \Delta_1 + \Gamma_{12} + \Gamma_{31} + \Gamma_{32})}\adop_1 \aop_1
    \\
    &\hspace{1em}+ \frac{4 \abs{g_2}^2 \Gamma_{12} (\Gamma_{21} + \Gamma_{31} + \Gamma_{32})}{(\Gamma_{12} + \Gamma_{21}) (\Gamma_{31} + \Gamma_{32}) (-2 i \Delta_2 + \Gamma_{21} + \Gamma_{31} + \Gamma_{32}) (2 i \Delta_2 + \Gamma_{21} + \Gamma_{31} + \Gamma_{32})} \adop_2 \aop_2,
\end{split}
\end{equation*}

\begin{equation}\begin{split}\label{eq:sigma11_2}
    \hat{\sigma}_{11} ^{(2)} = 
    &-\frac{4\abs{g_1}^2 \Gamma_{21} (\Gamma_{12} + \Gamma_{31} + \Gamma_{32}) (\Gamma_{21} + \Gamma_{32})}{(\Gamma_{12} + \Gamma_{21})^2 (\Gamma_{31} + \Gamma_{32}) (-2 i \Delta_1 + \Gamma_{12} + \Gamma_{31} + \Gamma_{32}) (2 i \Delta_1 + \Gamma_{12} + \Gamma_{31} + \Gamma_{32})} \adop_1\aop_1 \\
    &+\frac{4\abs{g_2}^2 \Gamma_{12} (-\Gamma_{21} + \Gamma_{31}) (\Gamma_{21} + \Gamma_{31} + \Gamma_{32})}{(\Gamma_{12} + \Gamma_{21})^2 (\Gamma_{31} + \Gamma_{32}) (-2 i \Delta_2 + \Gamma_{21} + \Gamma_{31} + \Gamma_{32}) (2 i \Delta_2 + \Gamma_{21} + \Gamma_{31} + \Gamma_{32})}\adop_2 \aop_2,
\end{split}\end{equation}

\begin{equation}\begin{split}\label{eq:sigma22_2}
    \hat{\sigma}_{22} ^{(2)} = 
    &- \frac{4\abs{g_2}^2 \Gamma_{12} (\Gamma_{12} + \Gamma_{31}) (\Gamma_{21} + \Gamma_{31} + \Gamma_{32})}{(\Gamma_{12} + \Gamma_{21})^2 (\Gamma_{31} + \Gamma_{32}) (-2 i \Delta_2 + \Gamma_{21} + \Gamma_{31} + \Gamma_{32}) (2 i \Delta_2 + \Gamma_{21} + \Gamma_{31} + \Gamma_{32})}\adop_2\aop_2 \\
    &+ \frac{4\abs{g_1}^2 \Gamma_{21} (\Gamma_{32} - \Gamma_{12}) (\Gamma_{12} + \Gamma_{31} + \Gamma_{32})}{(\Gamma_{12} + \Gamma_{21})^2 (\Gamma_{31} + \Gamma_{32}) (-2 i \Delta_1 + \Gamma_{12} + \Gamma_{31} + \Gamma_{32}) (2 i \Delta_1 + \Gamma_{12} + \Gamma_{31} + \Gamma_{32})}\adop_1\aop_1.
\end{split}\end{equation}

The non-zero terms at third order are given by:
\be\begin{split}
    \hat{\sigma}_{13} ^{(3)}
    =& -\frac{8i g_1\abs{g_1}^2 \Gamma_{21} (\Gamma_{12} + 2 \Gamma_{21} + \Gamma_{32}) (\Gamma_{12} + \Gamma_{31} + \Gamma_{32})}{(\Gamma_{12} + \Gamma_{21})^2 (\Gamma_{31} + \Gamma_{32}) (-2 i \Delta_1 + \Gamma_{12} + \Gamma_{31} + \Gamma_{32}) (2 i \Delta_1 + \Gamma_{12} + \Gamma_{31} + \Gamma_{32})^2}\adop_1\aop_1\aop_1
    + \frac{\Sigma_{13}}{\Xi_{13}}\adop_2\aop_2\aop_1,
\end{split}\ee
where $\Sigma_{13} = -8 g_1 \abs{g_2}^2 (-4 \Delta_2^2 \Gamma_{21} (\Gamma_{12} + \Gamma_{21}) (\Gamma_{31} + \Gamma_{32}) + 4 \Delta_1^2 \Gamma_{12} (\Gamma_{12} + 2 \Gamma_{21} - \Gamma_{31}) (\Gamma_{21} + \Gamma_{31} + \Gamma_{32}) + 2 i \Delta_2 \Gamma_{12} (\Gamma_{12} + \Gamma_{31} + \Gamma_{32}) (\Gamma_{12} \Gamma_{21} + 
2 \Gamma_{21}^2 + \Gamma_{21} \Gamma_{32} - \Gamma_{31} (\Gamma_{31} + \Gamma_{32})) - (\Gamma_{12} + \Gamma_{21}) (\Gamma_{21} + \Gamma_{31} + \Gamma_{32}) (\Gamma_{12}^3 + \Gamma_{12} (2 \Gamma_{21} + \Gamma_{32}) (\Gamma_{31} + \Gamma_{32}) + \Gamma_{21} (\Gamma_{31} + \Gamma_{32}) (\Gamma_{21} + \Gamma_{31} + \Gamma_{32}) + \Gamma_{12}^2 (2 \Gamma_{21} + \Gamma_{31} + 2 \Gamma_{32})) - 2 \Delta_1 \Gamma_{12} (2 \Delta_2 (\Gamma_{12} \Gamma_{21} + 2 \Gamma_{21}^2 + \Gamma_{21} \Gamma_{32} - \Gamma_{31} (\Gamma_{31} + \Gamma_{32})) + i (\Gamma_{21} + \Gamma_{31} + \Gamma_{32}) (2 \Gamma_{12}^2 + 2 \Gamma_{21}^2 - \Gamma_{31} (\Gamma_{31} + \Gamma_{32}) + \Gamma_{12} (5 \Gamma_{21} + 2 \Gamma_{32}) + \Gamma_{21} (2 \Gamma_{31} + 3 \Gamma_{32})))) $ and $\Xi_{13}=(\Gamma_{12} + \Gamma_{21})^2 (-2 \Delta_1 + 2 \Delta_2 + i (\Gamma_{12} + \Gamma_{21})) (\Gamma_{31} + \Gamma_{32}) (2 i \Delta_1 + \Gamma_{12} + \Gamma_{31} + \Gamma_{32})^2 (-2 i \Delta_2 + \Gamma_{21} + \Gamma_{31} + \Gamma_{32}) (2 i \Delta_2 + \Gamma_{21} + \Gamma_{31} + \Gamma_{32})$.
\be\begin{split}
    \hat{\sigma}_{23} ^{(3)} &= -\frac{8i g_2\abs{g_2}^2 \Gamma_{12} (2\Gamma_{12} + \Gamma_{21} + \Gamma_{31}) (\Gamma_{21} + \Gamma_{31} + \Gamma_{32})}{(\Gamma_{12} + \Gamma_{21})^2 (\Gamma_{31} + \Gamma_{32}) (-2 i \Delta_2 + \Gamma_{21} + \Gamma_{31} + \Gamma_{32}) (2 i \Delta_2 + \Gamma_{21} + \Gamma_{31} + \Gamma_{32})^2}\adop_2\aop_2\aop_2
    +\frac{\Sigma_{23}}{\Xi_{23}}\adop_1\aop_1\aop_2,
\end{split}\ee
where $\Sigma_{23} = 8 \abs{g_1}^2 g_2 (4 \Delta_1^2 \Gamma_{12} (\Gamma_{12} + \Gamma_{21}) (\Gamma_{31} + \Gamma_{32}) + 2 \Delta_1 \Gamma_{21} (2 \Gamma_{12}^2 + \Gamma_{12} (\Gamma_{21} + \Gamma_{31}) - \Gamma_{32} (\Gamma_{31} + \Gamma_{32})) (2 \Delta_2 - i (\Gamma_{21} + \Gamma_{31} + \Gamma_{32})) + (\Gamma_{12} + \Gamma_{31} + \Gamma_{32}) (-4 \Delta_2^2 \Gamma_{21} (2 \Gamma_{12} + \Gamma_{21} - \Gamma_{32}) + 2 i \Delta_2 \Gamma_{21} (2 \Gamma_{12}^2 + 2 \Gamma_{21}^2 + 2 \Gamma_{21} \Gamma_{31} - \Gamma_{32} (\Gamma_{31} + \Gamma_{32}) + \Gamma_{12} (5 \Gamma_{21} + 3 \Gamma_{31} + 2 \Gamma_{32})) + (\Gamma_{12} + \Gamma_{21}) (\Gamma_{12}^2 (\Gamma_{31} + \Gamma_{32}) + \Gamma_{21} (\Gamma_{21} + \Gamma_{31}) (\Gamma_{21} + \Gamma_{31} + \Gamma_{32}) + \Gamma_{12} (2 \Gamma_{21}^2 + 2 \Gamma_{21} (\Gamma_{31} + \Gamma_{32}) + (\Gamma_{31} + \Gamma_{32})^2)))) $ and $\Xi_{23}=(\Gamma_{12} + \Gamma_{21})^2 (2 \Delta_1 - 2 \Delta_2 + i (\Gamma_{12} + \Gamma_{21})) (\Gamma_{31} + \Gamma_{32}) (-2 i \Delta_1 + \Gamma_{12} + \Gamma_{31} + \Gamma_{32}) (2 i \Delta_1 + \Gamma_{12} + \Gamma_{31} + \Gamma_{32}) (2 i \Delta_2 + \Gamma_{21} + \Gamma_{31} + \Gamma_{32})^2$.
\be\begin{split}
    \hat{\sigma}_{31} ^{(3)} &= \frac{8 i g_1^*\abs{g_1}^2 \Gamma_{21} (\Gamma_{12} + 2 \Gamma_{21} + \Gamma_{32}) (\Gamma_{12} + \Gamma_{31} + \Gamma_{32}) }{(\Gamma_{12} + \Gamma_{21})^2 (\Gamma_{31} + \Gamma_{32}) (-2 i \Delta_1 + \Gamma_{12} + \Gamma_{31} + \Gamma_{32})^2 (2 i \Delta_1 + \Gamma_{12} + \Gamma_{31} + \Gamma_{32})}\adop_1\adop_1\aop_1 + \frac{\Sigma_{31}}{\Xi_{31}}\adop_1\adop_2\aop_2,
\end{split}\ee
where $\Sigma_{31}=-8 g_1^*\abs{g_2}^2 (-4 \Delta_2^2 \Gamma_{21} (\Gamma_{12} + \Gamma_{21}) (\Gamma_{31} + \Gamma_{32}) + 4 \Delta_1^2 \Gamma_{12} (\Gamma_{12} + 2 \Gamma_{21} - \Gamma_{31}) (\Gamma_{21} + \Gamma_{31} + \Gamma_{32}) - 2 i \Delta_2 \Gamma_{12} (\Gamma_{12} + \Gamma_{31} + \Gamma_{32}) (\Gamma_{12} \Gamma_{21} + 2 \Gamma_{21}^2 + \Gamma_{21} \Gamma_{32} - \Gamma_{31} (\Gamma_{31} + \Gamma_{32})) - (\Gamma_{12} + \Gamma_{21}) (\Gamma_{21} + \Gamma_{31} + \Gamma_{32}) (\Gamma_{12}^3 + \Gamma_{12} (2 \Gamma_{21} + \Gamma_{32}) (\Gamma_{31} + \Gamma_{32}) + \Gamma_{21} (\Gamma_{31} + \Gamma_{32}) (\Gamma_{21} + \Gamma_{31} + \Gamma_{32}) + \Gamma_{12}^2 (2 \Gamma_{21} + \Gamma_{31} + 2 \Gamma_{32})) + \Delta_1 (-4 \Delta_2 \Gamma_{12} (\Gamma_{12} \Gamma_{21} + 2 \Gamma_{21}^2 + \Gamma_{21} \Gamma_{32} - \Gamma_{31} (\Gamma_{31} + \Gamma_{32})) + 2 i \Gamma_{12} (\Gamma_{21} + \Gamma_{31} + \Gamma_{32}) (2 \Gamma_{12}^2 + 2 \Gamma_{21}^2 - \Gamma_{31} (\Gamma_{31} + \Gamma_{32}) + \Gamma_{12} (5 \Gamma_{21} + 2 \Gamma_{32}) + \Gamma_{21} (2 \Gamma_{31} + 3 \Gamma_{32})))) $ and $\Xi_{31} =(\Gamma_{12} + \Gamma_{21})^2 (2 \Delta_1 - 2 \Delta_2 + i (\Gamma_{12} + \Gamma_{21})) (\Gamma_{31} + \Gamma_{32}) (-2 i \Delta_2 + \Gamma_{21} + \Gamma_{31} + \Gamma_{32}) (2 i \Delta_2 + \Gamma_{21} + \Gamma_{31} + \Gamma_{32}) (2 \Delta_1 + i (\Gamma_{12} + \Gamma_{31} + \Gamma_{32}))^2$.
\be\begin{split}
    \hat{\sigma}_{32} ^{(3)} &= \frac{8 i g_2^*\abs{g_2}^2 \Gamma_{12} (2 \Gamma_{12} + \Gamma_{21} + \Gamma_{31}) (\Gamma_{21} + \Gamma_{31} + \Gamma_{32})}{(\Gamma_{12} + \Gamma_{21})^2 (\Gamma_{31} + \Gamma_{32}) (-2 i \Delta_2 + \Gamma_{21} + \Gamma_{31} + \Gamma_{32})^2 (2 i \Delta_2 + \Gamma_{21} + \Gamma_{31} + \Gamma_{32})}\adop_2\adop_2\aop_2
    +\frac{\Sigma_{32}}{\Xi_{32}}\adop_2\adop_1\aop_1,
\end{split}\ee
where $\Sigma_{32} = 8 \abs{g_1}^2g_2^* (4 \Delta_1^2 \Gamma_{12} (\Gamma_{12} + \Gamma_{21}) (\Gamma_{31} + \Gamma_{32}) + 2 \Delta_1 \Gamma_{21} (2 \Gamma_{12}^2 + \Gamma_{12} (\Gamma_{21} + \Gamma_{31}) - \Gamma_{32} (\Gamma_{31} + \Gamma_{32})) (2 \Delta_2 + i (\Gamma_{21} + \Gamma_{31} + \Gamma_{32})) + (\Gamma_{12} + \Gamma_{31} + 
\Gamma_{32}) (-4 \Delta_2^2 \Gamma_{21} (2 \Gamma_{12} + \Gamma_{21} - \Gamma_{32}) - 2 i \Delta_2 \Gamma_{21} (2 \Gamma_{12}^2 + 2 \Gamma_{21}^2 + 2 \Gamma_{21} \Gamma_{31} - \Gamma_{32} (\Gamma_{31} + \Gamma_{32}) + \Gamma_{12} (5 \Gamma_{21} + 3 \Gamma_{31} + 2 \Gamma_{32})) + (\Gamma_{12} + \Gamma_{21}) (\Gamma_{12}^2 (\Gamma_{31} + \Gamma_{32}) + \Gamma_{21} (\Gamma_{21} + \Gamma_{31}) (\Gamma_{21} + \Gamma_{31} + \Gamma_{32}) + \Gamma_{12} (2 \Gamma_{21}^2 + 2 \Gamma_{21} (\Gamma_{31} + \Gamma_{32}) + (\Gamma_{31} + \Gamma_{32})^2))))$ and $\Xi_{32} = (\Gamma_{12} + \Gamma_{21})^2 (-2 \Delta_1 + 2 \Delta_2 + i (\Gamma_{12} + \Gamma_{21})) (\Gamma_{31} + \Gamma_{32}) (-2 i \Delta_1 + \Gamma_{12} + \Gamma_{31} + \Gamma_{32}) (2 i \Delta_1 + \Gamma_{12} + \Gamma_{31} + \Gamma_{32}) (2 \Delta_2 + i (\Gamma_{21} + \Gamma_{31} + \Gamma_{32}))^2$.

\section{Microscopic quantization for the case of distinguishable fields}\label{app:difffields}

The microscopic quantization of this system is still described by the single-excitation eigenstates of \eqnref{eq:Heff}. The single-excitation subspace is now spanned by $\cpare{\adop_1\ket{1}, \adop_2\ket{2},\ket{3}}$. One can again obtain two types of polariton operators, 
$\hat{A}^\dag_j$ and $\hat{B}^\dag_j$, corresponding to creating an excitation by their respective action on the zero-excitation states $\ket{1}$ and $\ket{2}$.
They have the form:
\be
    \hat{A}_j^\dag = c_{1j} \hat{a}^\dag_1\hat{\sigma}_{11} + c_{2j} \hat{a}^\dag_2\hat{\sigma}_{21} + c_{3j} \hat{\sigma}_{31},
\ee
\be
    \hat{B}_j^\dag = c_{1j} \hat{a}^\dag_1\hat{\sigma}_{12} + c_{2j} \hat{a}^\dag_2\hat{\sigma}_{22} + c_{3j} \hat{\sigma}_{32}.
\ee
Note that this is different from the single field case in two ways. First, now there are two kinds of photonic contributions, of types $\adop_j$, reflecting the fact that there are two fields. Second, the coefficients depend on $\Gamma_{12}$ and $\Gamma_{21}$ which were both set to zero earlier. In particular, they are given by
\be
    c_{1j} = \frac{1}{\sqrt{1+\frac{k_j^2+\pare{\kappa_j+\Gamma_{12}}^2}{g_1^2} + \frac{4g_2^2\pare{k_j^2+\pare{\kappa_j+\Gamma_{12}}^2}}{g_1^2\pare{4 (k_j + \Delta_1 - \Delta_2)^2 + (\Gamma_{12} + \Gamma_{21} + 2 \kappa_j)^2}}}},
\ee
\be
    c_{2j} = \frac{2g_2\pare{\kappa_j+\Gamma_{12}-ik_j}}{g_1\pare{\Gamma_{12} + \Gamma_{21}-2i\pare{k_j+\Delta_1-\Delta_2}+2\kappa_j}\sqrt{1+\frac{k_j^2+\pare{\kappa_j+\Gamma_{12}}^2}{g_1^2} + \frac{4g_2^2\pare{k_j^2+\pare{\kappa_j+\Gamma_{12}}^2}}{g_1^2\pare{4 (k_j + \Delta_1 - \Delta_2)^2 + (\Gamma_{12} + \Gamma_{21} + 2 \kappa_j)^2}}}},
\ee
\be
    c_{3j} =  \frac{k_j+i\kappa_j+i\Gamma_{12}}{g_1\sqrt{1+\frac{k_j^2+\pare{\kappa_j+\Gamma_{12}}^2}{g_1^2} + \frac{4g_2^2\pare{k_j^2+\pare{\kappa_j+\Gamma_{12}}^2}}{g_1^2\pare{4 (k_j + \Delta_1 - \Delta_2)^2 + (\Gamma_{12} + \Gamma_{21} + 2 \kappa_j)^2}}}},
\ee
where $k_j\equiv\Re\pare{E_j}$ and $\kappa_j\equiv\Im\pare{E_j}$. As in the single field case, the overall phase of the coefficients $c_{ij}$ is chosen such that $c_{1j} \in \mathbb{R}$ and positive. The additional loss terms also modify the deviation of the polariton operators from canonical bosonic commutation relations which is now given by $\langle 1|\left[\hat{A}_i,\hat{A}_j^\dag\right]|1 \rangle = \langle 2|\left[\hat{B}_i,\hat{B}_j^\dag\right]|2 \rangle = \delta_{ij}+\pare{1-\delta_{ij}} i\pare{\Gamma_{12} c_{1i}^*c_{1j} + \Gamma_{21} c_{2i}^*c_{2j}+\pare{\Gamma_{31}+\Gamma_{32}} c_{3i}^*c_{3j}}/\pare{\pare{E_i}^*-E_j}$.

\end{document}